\documentclass[aps,pra,reprint,superscriptaddress,footnote, twocolumn,notitlepage,longbibliography]{revtex4-1}
\usepackage{times,bbm}
\usepackage[usenames,dvipsnames]{xcolor}
\usepackage{mathtools}
\usepackage[colorlinks=true,linkcolor=blue,urlcolor=blue,citecolor=blue]{hyperref}
\usepackage{graphics}
\usepackage{bm}
\usepackage{color}
\usepackage{amscd}
\usepackage{amsfonts}
\usepackage{amssymb}
\usepackage{amsmath}
\usepackage{bbold}
\usepackage{graphicx}
\usepackage{tabularx}
\usepackage{braket}
\usepackage{bbm}

\newcommand{\ketbra}[2]{|#1\rangle\langle#2|}

\begin{document}

\title{Nonequilibrium steady states in multi-bath quantum collision models} 

\author{Ronan McElvogue}
\affiliation{School of Physics, University College Dublin, Belfield, Dublin 4, Ireland}
\affiliation{Centre for Quantum Engineering, Science, and Technology, University College Dublin, Dublin 4, Ireland}

\author{Andrew K. Mitchell}
\affiliation{School of Physics, University College Dublin, Belfield, Dublin 4, Ireland}
\affiliation{Centre for Quantum Engineering, Science, and Technology, University College Dublin, Dublin 4, Ireland}

\author{Gabriel T. Landi}
\affiliation{Department of Physics and Astronomy, University of Rochester, Rochester, New York 14627, USA}
\affiliation{University of Rochester Center for Coherence and Quantum Science, Rochester, New York 14627, USA}

\author{Steve Campbell}
\affiliation{School of Physics, University College Dublin, Belfield, Dublin 4, Ireland}
\affiliation{Centre for Quantum Engineering, Science, and Technology, University College Dublin, Dublin 4, Ireland}

\begin{abstract}
Collision models provide a simple and versatile setting to capture the dynamics of open quantum systems. The standard approach to thermalisition in this setting involves an environment of independent and identically-prepared thermal qubits, interacting sequentially for a finite duration $\Delta t$ with the system. We compare this to a two-bath scenario in which collisional qubits are prepared in either their ground or excited states and the environment temperature is encoded in system-environment couplings. The system reaches the same thermal steady state for both settings, although even in this limit they describe fundamentally different physical processes, with the two-bath setup yielding a \textit{nonequilibrium} state with finite heat currents. Non-Markovian dynamics arise when intra-environment interactions in either setting are introduced. Here, the system in the single-bath setup again reaches a steady state at the canonical temperature of the bath, but the nonequilibrium steady state of the two-bath setup tends to a different temperature due to the generation of strong system-environment and intra-environment correlations. The two-bath setting is particularly suited to studying quantum trajectories, which are well-defined also for the non-Markovian case. We showcase this with a trajectory analysis of the heat currents within a two-point measurement scheme. Finally, we consider how our results are impacted when the system-environment interaction leads to strict homogenisation. Our results provide insights into the dynamics and thermodynamics of thermalisation towards nonequilibrium steady states and the role of non-Markovian interactions.
\end{abstract}

\maketitle

\section{Introduction}
A framework for understanding how a quantum system evolves when in contact with an environment is critical; both in terms of the practical development of quantum technologies and in addressing long-standing open questions regarding the dynamics of complex systems. Regarding the former noise characterisation in prototype quantum devices is crucial to overcome scalability issues~\cite{ManiscalcoScalability}, while the latter touches on major conceptual gaps in our understanding, e.g. determining the quantum-to-classical transition~\cite{ZurekRMP} or how complex systems thermalise~\cite{Deutsch2018}. It is not surprising then that a range of tools have been developed to simulate open quantum systems~\cite{PReB, Ferracin2024, LandiRMP, Breuer2002, GooldPRX}. Within this plethora of techniques, collision models~\cite{CICCARELLO20221, Campbell2021EPL, cusumano2022quantum, Rau1963, ScaraniPRL2002, Homogenisation} or repeated interaction schemes~\cite{StrasbergPRX}, stand as a particularly versatile approach having been used to reconcile apparent thermodynamic inconsistencies present in other approaches~\cite{DeChiaraNJP}, explore how classicality emerges from an otherwise quantum dynamics~\cite{CampbellPRA2019, Ryan2022}, explore synchronization in quantum systems~\cite{karpat2019pra}, and efficiently cool quantum systems~\cite{HuberCooling, KochCooling}. 

The simplest formulation of a collision model involves imagining the environment as an infinite collection of identically prepared constituents, here after referred to as ``auxiliary units'' and labelled $A$, with which the system, $S$, sequentially interacts. Depending on the precise form of the $S$-$A$ interaction, and the initial states of the auxiliary units, this model accurately captures a range of Markovian dynamics. In particular, for so-called ``thermal operations'' it leads to thermalisation~\cite{ScaraniPRL2002}, while the model also more generally leads to homogenisation~\cite{Homogenisation}. However, varying the $S$-$A$ interaction can lead to a range of non-equilibrium steady states (NESS), whose characterisation has recently become the focus of several works~\cite{Guarnieri2020PLA, Heineken_NESS_Entanglement, Tian_Heat_Transfer_2021, Segal2025QST, Baris1, Baris2, Hammam2024quantum, cusumano2024structured, hammam2021optimizing, corr2025continuous}. While assuming that all the auxiliary units are identically prepared is well justified, particularly in order to simplify simulation, this is clearly not a very realistic model. For instance, a more precise microscopic description of a thermal environment would assume that the system interacts with auxiliary units whose initial state is drawn in accordance with a distribution of energy that correspond to a given temperature~\cite{KochCooling}. Several works have explored relaxing the assumption of identical auxiliary units and deterministic collisional events, e.g. by leveraging scattering techniques~\cite{Scattering1, Scattering2, Scattering3}. The flexibility of collision models is clearly one of their inherent strengths that allows us to explore a range of other open system dynamics~\cite{Lorenzo2015heat, cusumano2017interferometric, cusumano2018interferometric, pedram2022environment, Erbanni2023simulating, Garg2025simulating}. From the preceding discussion it is clear that collision models provide a useful approach to simulate how a system reaches the same, possibly thermal or equilibrium, steady state from different microscopic models.

This serves as the starting point for our analysis. Our aim is to critically explore the differences between collision models involving a single thermal bath and those featuring multiple pure-state baths, particularly in the non-Markovian regime. We start by considering a simple thermalisation process via collision models for two settings: the `typical' situation where the auxiliary units are assumed to be all identical thermal states and a second setting where the system interacts simultaneously with two collisional baths with different initial energies, and these energies dictate the strength of the $S$-$A$ interaction. While these approaches are known to recover the expected Markovian dynamics as described by the standard GKSL master equation~\cite{CICCARELLO20221, cusumano2022quantum} (see the Appendix), we show that they correspond to physically distinct processes, with the latter setting clearly driving the system to a NESS. Through a simple modification of the basic collision model, we examine how the introduction of non-Markovianity to the dynamics significantly changes the steady state properties observed, with the establishment of strong system bath correlations playing a particularly important role. We explore the impact of introducing projective measurements on the auxiliary units using a two-point measurement scheme to recover the average heat current in the presence of non-Markovianity and briefly comment on the impact that alternative system-environment interaction terms can have on the steady-state properties.

\section{Markovian Quantum Collision models}
\label{sec:Markovian} 

\subsection{Single-bath thermalizing model (setting I)}

We consider the typical collision model set-up which consists of a system of interest $S$, in an initial state $\rho_{0}$, interacting with an environment which is partitioned into individual constituents, the ``auxiliary units''. These auxiliary units, labelled $A_n$, are each initialised in the state $\eta_n$. The time evolution is modeled in a stroboscopic manner by allowing the auxiliary units to `collide' with the system for a fixed time $\Delta t$, such that a single interaction is given by the unitary operator (here and henceforth we set $\hbar =k_B = 1$),
\begin{equation}
\label{eq:unitary}
U_n = e^{-i(H_S + H_{A_n} + H_{I})\Delta t} \;,
\end{equation}
where $H_S$ and $H_{A_n}$ are the free Hamiltonians of the system and auxiliary unit, respectively, and $H_{I}$ is the interaction Hamiltonian between $S$ and $A_n$. After $A_n$ has completed its interactions, we trace over its degrees of freedom. This procedure leads to the open system dynamics of $S$ described by the map,
\begin{equation}
\rho_n = \Phi [\rho_0]=\mathcal{E}^{n} \circ \mathcal{E}^{n-1}\cdots \circ \mathcal{E}^{1}\left[\rho_0 \right] \;, 
\end{equation}
where $\circ$ indicates concatenation and,
\begin{equation} \label{Collision Map}
    \mathcal{E}^j[\rho_{j-1}]= \text{Tr}_j[U_j(\rho_{j-1} \otimes \eta_j)U^{\dag}_j] \;,
\end{equation}
is the quantum channel describing the collision with the $j$-th auxiliary unit. 

\begin{figure}[t]
    \includegraphics[width=1\linewidth]{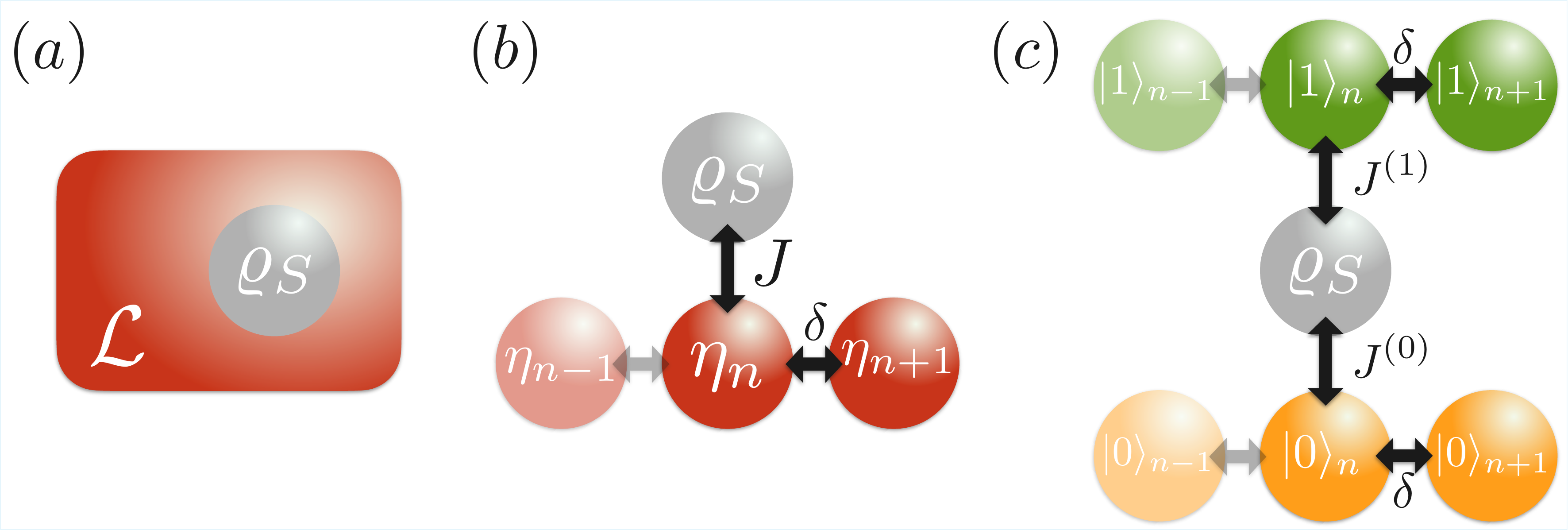}
    \caption{Schematic depiction of the approaches to model open quantum system dynamics considered in this work. (a) Typical open system dynamics where the system is interacts with a bath, generally assumed to be much larger than the system. Under certain assumptions this can be modeled using master equations. (b) Setting I: Collision models involve partitioning the environment into smaller constituent parts all prepared in the same initial state, $\eta_i$. The system interacts with a given environmental unit via some interaction with strength $J$. For interactions of the form Eq.~\eqref{eqn: interaction H} this collision model leads to homogenisation and the system reaches the steady state $\rho_{\infty}\!=\!\eta_i$. If $\eta_i$ are thermal states, this dynamics can be shown to be equivalent to the master equation, Eq.~\eqref{Eqn: QME}. (c) Setting II: Two bath collision model. The system interacts simulataneously with two independent collisional baths, where the constituents are initially in pure states $\ket{0}$ and $\ket{1}$, respectively, whose interaction strengths encode the effective temperature of the bath.}
    \label{fig:NM CM Sketch}
\end{figure}

Collision models are a flexible framework for studying open system dynamics~\cite{CICCARELLO20221, Campbell2021EPL}; indeed they can be used to simulate any time-continuous open system dynamics described by a Markovian quantum master equation~\cite{MEsfromCMs, Cattaneo2021collision}. A simple setting of wide physical relevance is a two-level system with Hamiltonian, $H_S=\tfrac{\omega_S}{2}\sigma_z$, thermalising with a bosonic environment described by the master equation~\cite{Current_fluctuations},
\begin{equation}\label{Eqn: QME}
    \dfrac{d\rho}{dt} = \mathcal{L}(\rho) = -i \left[H_S,\rho \right]+\Gamma(\bar{N}+1)\mathcal{D}[\sigma^-]\rho +\Gamma\bar{N}\mathcal{D}[\sigma^+]\rho \;,
\end{equation}
where $\Gamma$ is the coupling strength and $\bar{N}\!=\! (e^{\beta \omega_S}-1)^{-1}$ is the Bose-Einstein distribution for a bath at (inverse) temperature $\beta\!=\!1/T$, as depicted in Fig.~\ref{fig:NM CM Sketch}(a). The two dissipators $\mathcal{D}[\sigma^-], \mathcal{D}[\sigma^+]$ represent the emission and absorption of thermal photons, respectively, with $\sigma^-\!=\!\ketbra{0}{1}$ and $\sigma^+\!=\!\ketbra{1}{0}$ being the Pauli lowering and raising operators. The steady state of Eq.~\eqref{Eqn: QME} is given by the Gibbs state,
\begin{equation}
\label{eq:Gibbsstate}
\rho^\text{th} = \begin{pmatrix} \frac{1-g}{2} & 0 \\ 0 & \frac{1+g}{2} \end{pmatrix}.
\end{equation}
where $g = (2\Bar{N}+1)^{-1}$~\cite{2-level_thermalization}. 

The dynamics described by Eq.~\eqref{Eqn: QME} can be stroboscopically simulated using a collision model~\cite{Rau1963, ScaraniPRL2002, CICCARELLO20221, cusumano2022quantum}, where the environment is viewed as an infinite collection of auxiliary units, each modeled as a qubit, identically prepared in a thermal state, i.e. $\eta_n=\rho^\text{th}$ given by Eq.~\eqref{eq:Gibbsstate}, with the free evolution of each unit given by $H_{A_{n}}=\tfrac{\omega_{A_{n}}}{2} \sigma^z$. For simplicity, in the rest of this work we shall always assume the system and auxiliary units are on resonance, i.e. $\omega_S \!=\! \omega_{A_{n}} \!\equiv\! \omega$. This setting is illustrated in Fig.~\ref{fig:NM CM Sketch}(b). In order to match the dynamics given by Eq.~\eqref{Eqn: QME}, we fix the interaction Hamiltonian between the $S$ and $A_n$ to be,
\begin{equation}\label{eqn: interaction H}
    H_{I} = -\frac{J}{2}\left(\sigma^x_S \otimes \sigma^x_{A_n} + \sigma^y_S \otimes \sigma^y_{A_n}\right) \;,
\end{equation}
with $J = \sqrt{{\frac{\Gamma(2\bar{N}+1)}{\Delta t}}}$.
Scaling the interaction strength by $1/\sqrt{\Delta t}$ is necessary to ensure that Eq.~\eqref{Eqn: QME} is recovered in the limit $\Delta t\to0$.
Throughout this work we will refer to this as `setting I'. Henceforth, we shall move to the interaction picture and neglect the free evolution of the system and auxiliary units. This can be done due to the fact that sum of the free Hamiltonians commute with the interaction Hamiltonian, i.e. $\left [H_S+H_{A_{n}},H_I\right] = 0$ ~\cite{Composite_CM}. This fact also means there is no switching work required to stoboscopically turn on and off the system-bath interaction~\cite{DeChiaraNJP, Guarnieri2020PLA}.

This setup leads to thermalization, wherein the steady state of $S$ converges to the initial auxiliary state, i.e. $\rho_\infty\! =\! \eta_i$, for any choice of initial state for the auxiliary units $A_n$~\cite{Homogenisation}. When $S$ sequentially interacts with $A_n$ via Eq.~\eqref{eqn: interaction H}, its dynamics are described by a contractive completely positive map, guaranteeing convergence toward $\eta$ as the unique fixed point. This process is thermodynamically consistent and captures key features of thermalization. Importantly, in the limit of infinitely many collisions, the system not only reaches equilibrium but also asymptotically factorizes from the environment, resulting in a product state without residual system-environment correlations \cite{GiovannettiPRA2018}.

While here we will focus on system-environment interactions of the form Eq.~\eqref{eqn: interaction H}, in the Appendix we consider an alternative given by the isotropic Heisenberg Hamiltonian. Although this interaction also leads to homogenization in setting I, the collision model no longer reduces to the same thermalisation dynamics governed by the GKSL equation, Eq.~\eqref{Eqn: QME}~\cite{cusumano2024structured}. Indeed, while many of the main features discussed below are similarly exhibited for the Heisenberg interaction, we highlight a number of interesting differences that emerge when alternative methods for simulating the open dynamics are considered in both the Markovian and non-Markovian regimes.

\subsection{Two-bath model (setting II)}
\label{twobathsubsec}
An alternative approach to simulating the time-continuous dynamics of Eq.~\eqref{Eqn: QME} involves allowing $S$ to interact with two collisional baths, one in which all the auxiliary units are in the $\ket{0}$ state and one in which they are all in the $\ket{1}$ state, where $\ket{0}$ and $\ket{1}$ corresponds to the ground and excited states of $A_n$, respectively. This setting is illustrated in Fig. \ref{fig:NM CM Sketch}(c). Throughout this work we will refer to this as `setting II'. The system interacts with a qubit from each bath simultaneously via an interaction Hamiltonian $H_I^{(i)}$ during a given collision, where we have introduced the superscript $i = 0,1$ to label interactions with the baths $\ket{0}$ and $\ket{1}$, respectively. The interaction Hamiltonians take the same form as Eq.~\eqref{eqn: interaction H} but with different interaction strengths, $J^{(i)}$, which are given by
\begin{equation}
\label{eq:TwoBathInts}
     J^{(0)} = \sqrt{\frac{\Gamma(\bar{N}+1)}{\Delta t}}, \quad J^{(1)} = \sqrt{\frac{\Gamma \bar{N}}{\Delta t}}
\end{equation}
and encode the expected (effective) bath temperature. The unitary for a single step in this approach that enters into Eq.~\eqref{eq:unitary} is therefore (already in the interaction picture)
\begin{equation} \label{eqn: Multiple Baths CM Unitary}
     U_n^{\text{II}} = e^{-i(H_{I}^{(0)}+H_{I}^{(1)})\Delta t}.
\end{equation}

In contrast to setting I, in this case the system will be driven to a non-equilibrium steady state (NESS), due to the simultaneous interaction with multiple inequivalent baths.
This is because, irrespective of whether the system thermalizes or not, there will always be a constant flow of energy from the $\ket{1}$ bath to the $\ket{0}$ bath. Similar to setting I however, we have $\left [H_S+H_{A_{n}},H_{I}^{(0)}+H_{I}^{(1)}\right] = 0$ and so the free evolutions of the system and auxiliary units can again be neglected.


\subsection{Steady States in the Markovian Limit}
\label{sec:Markovian}
We start our analysis by noting that settings I and II both simulate the dynamics of a system in contact with a thermal bath. Indeed, at the level of the system only, the two settings exhibit the same dynamics and steady state that would be expected under the time continuous evolution given by Eq.~\eqref{Eqn: QME}. Furthermore, it is evident that they are equivalent in the zero temperature limit, $\beta \rightarrow \infty$. This is an obvious consequence of the fact that, in this limit, $J^{(1)}\!\to\!0$ and thus the two bath case reduces to the zero temperature limit of setting I. However, as alluded to previously, and as we will show explicitly below, setting II generally drives the system to a NESS.

\begin{figure}[t]
\includegraphics[width=0.99\linewidth]{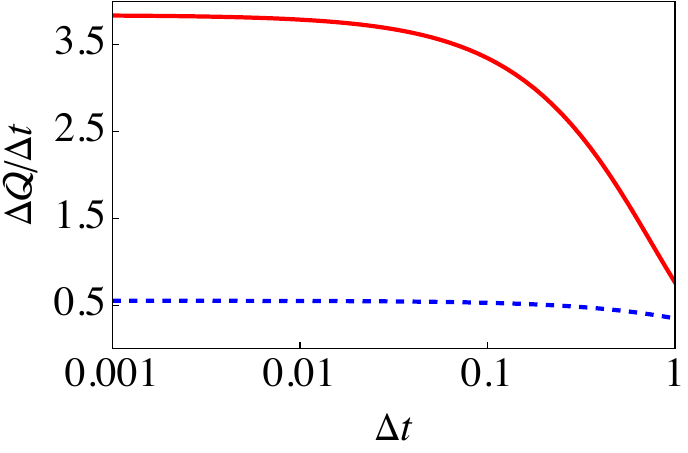}
\caption{Steady states in the Markovian limit. Non-equilibrium steady state heat current, $\Delta\mathcal{Q}/\Delta t$ for the initially low energy (cold) bath, $A_n^{(0)}$, with increasing collision time-step $\Delta t$ for different canonical temperatures $\beta\!=\!2$ (dashed, blue; lower temperature) and $\beta\!=\!0.5$ (solid, red; higher temperature). Other parameters: $\omega\!=\!1,~\Gamma\!=\!4$.}
\label{fig:Markovian}
\end{figure}

Maintaining the system in a NESS requires heat currents to be flowing. Because the unitary is strictly energy preserving, there is no work associated with switching on and off the collisional unitaries~\cite{DeChiaraNJP, Guarnieri2020PLA}. Hence all energy changes can be considered as heat exchanges. Focusing on the zero temperature bath in setting II, within the collision model setting we can readily calculate this via
\begin{equation}
\label{eq:Heat}
\Delta \mathcal{Q}^{(0)}(n) = \text{Tr}\left[ H_{A_n^{(0)}} \left(\tilde{\rho}_{A_n^{(0)}} - \rho_{A_n^{(0)}} \right) \right]
\end{equation}
where $\tilde{\rho}_{A_n^{(0)}}$ is the state of the $A_{n}^{(0)}$ auxiliary unit after the system-environment collision has taken place. A similar expression holds for the $A_{n}^{(1)}$'s and, in fact, we find is exactly equal and opposite to Eq.~\eqref{eq:Heat}, which is a direct consequence of strict energy conservation. Fig.~\ref{fig:Markovian} shows the steady state heat flux into the environment, i.e. $\frac{\Delta\mathcal{Q}}{\Delta t}\!\equiv\!\frac{\Delta\mathcal{Q}^{(0)}(n)}{\Delta t}$ for  $n\!\rightarrow\! \infty$. We see that it is non-zero for all values of $\Delta t$. Smaller values of $\beta$ require larger heat fluxes to maintain the NESS. This is in line with expectation since in this regime the system is being driven to a state with a high effective temperature. It is therefore far from equilibrium from both of the respective collisional baths leading to substantial energy exchanges. Larger $\beta$ leads to a reduced heat flux since this regime tends to the zero temperature single bath situation captured by setting I. It is worth noting that the heat flux vanishes for all choices of $\beta$ in setting~I~\cite{Guarnieri2020PLA}. 

\section{Non-Markovian Quantum Collision Models}
\label{sec:nonMarkovian}
When using collision models to simulate Markovian dynamics, we make three assumptions regarding the constituents of the bath:  {\it (i)} they do not interact with each other; {\it (ii)} they are initially uncorrelated; {\it (iii)} each constituent interacts with $S$ only once. Relaxing these assumptions allows us to extend collision models to simulate non-Markovian dynamics~\cite{ruariMauro, csenyacsa2022entropy, BarisPRA2017, Pezzutto_Paternostro_2016, Zhang2019, Pleasance_2025}. 

Here, we introduce non-Markovianity by breaking assumption {\it (i)} and introducing intra-environment collisions between consecutive units $A_n$ and $A_{n+1}$. We assume that this interaction occurs after the system has collided with $A_n$ but before it comes into contact with $A_{n+1}$. In all instances we fix this interaction to be
\begin{equation}
\label{eq:NMinteraction}
    U_{AA} = \text{cos}(\delta)\mathbbm{1} - i \text{sin}(\delta)\mathcal{W}_{A_{n},A_{n+1}}
\end{equation}
where $\mathcal{W}$ is the SWAP operator,
\begin{equation}
\label{eq:SWAP}
\mathcal{W} = \begin{pmatrix}
1 & 0 & 0 & 0 \\
0 & 0 & 1 & 0 \\
0 & 1 & 0 & 0 \\
0 & 0 & 0 & 1 \\
\end{pmatrix},
\end{equation} 
and $0\!\leq\! \delta \!  < \! \frac{\pi}{2}$. For the two settings considered previously, the unitary implementing the map is then
\begin{equation}
\label{eq:NonMunitaries}
    U_{n} = 
    \begin{cases}
    U_{AA}(\delta)U^\text{I}_{n}, & \text{Single~bath~setting~I}\\
    U_{AA}^{(0)}(\delta)U_{AA}^{(1)}(\delta)U^\text{II}_{n}, & \text{Multiple~bath~setting~II}.
    \end{cases}
\end{equation}
As before, the additional superscripts in setting II  denote in which collisional bath the unitary acts. A sketch of these approaches is captured in Fig.~\ref{fig:NM CM Sketch}(b) and (c) where the difference compared with the Markovian case is the inclusion of intra-environment collisions with some non-zero value of $\delta$. 

The impact of intra-environment interactions is to correlate $S$ with $A_{n+1}$ before they have directly interacted with each other. This leads to a non-Markovian dynamics~\cite{NmReview} with the parameter $\delta$ controlling the strength of the non-Markovianity~\cite{ruariMauro}. For $\delta\!=\!0$ the dynamics reverts to the Markovian case as discussed in Sec.~\ref{sec:Markovian}, while $\delta\!=\! \frac{\pi}{2}$ leads to an infinitely non-Markovian dynamics~\cite{ruariMauro} since, for this interaction strength, Eq.~\eqref{eq:NMinteraction} becomes a full SWAP and therefore the system is effectively interacting with the same qubit in the environment at every timestep and there is no loss of information.

A useful way to understand the introduction of the intra-environment interactions is through the notion of a {\it memory-depth}~\cite{Bassano}. For the Markovian collision models in Sec.~\ref{sec:Markovian}, each environmental constituent is involved in only a single collision and, thus, can be traced out after this interaction has taken place since it plays no role in the ensuing dynamics. Clearly, this is not the case for the non-Markovian setting as each $A_n$ is involved in two collisions, one with the system and a second with the incoming bath qubit, $A_{n+1}$. Introducing an additional full SWAP operation applied to $A_n$ and $A_{n+1}$ in Eqs.~\eqref{eq:NonMunitaries} allows to recover an effective Markovian dynamics~\cite{Bassano, AA_to_Composite}. This is because the additional SWAP operation means that, in effect, the system interacts with the {\it same} bath unit, now considered a memory, during every collision. While this may seem like an unnecessary complication, it means that at the level of the system plus memory compound the dynamics is Markovian~\cite{Bassano, Composite_CM, AA_to_Composite}. This Markovian embedding of non-Markovian dynamics allows to readily determine the system-memory steady state, from which the steady state of the system directly follows by tracing over the memory's degrees of freedom.

Our interest is in characterising the steady states arising from the different microscopic models captured by Fig.~\ref{fig:NM CM Sketch}. The ability to introduce non-Markovianity via intra-environment interactions therefore adds an interesting additional dimension to our study, complementing recent works that focused on the Markovian limit~\cite{Guarnieri2020PLA, Segal2025QST}. To this end, we aim to assess the impact, if any, that degree of non-Markovianity plays in the resulting steady state properties. Several metrics have been proposed to quantify non-Markovianity~\cite{NmReview, NMmeasures}. We shall focus on the BLP measure~\cite{BLPmeasure, BLPmeasure2}, which takes an information theoretic approach based on the behaviour of the trace distance between two states,
\begin{equation}
\label{eq:tracedistance}
    D(\rho, \pi) = \frac{1}{2}||\rho - \pi||.
\end{equation}
Equation~\eqref{eq:tracedistance} is a quantifier of the distinguishability of two states, ranging from 1 for fully distinguishable states, to 0 for fully indistinguishable states. For a time continuous dynamics, the BLP measure is defined as
\begin{equation}
\label{eq:continous BLP}
    N = \max_{\{\rho(0),\pi(0)\}} \int_{\Omega_+}\partial_t D(\rho(t), \pi(t))dt ,
\end{equation}
where $\Omega_+ = \bigcup_i(a_i,b_i)$ is the union of all the time intervals in the dynamics in which the trace distance between $\rho$ and $\pi$ increases, i.e. $\partial_t D(\rho(t), \pi(t))>0$. Here, $\rho(t)$ and $\pi(t)$ start from initially distinguishable (orthogonal) states. Equation~\eqref{eq:continous BLP} can be interpreted as summing over all the regions where the trace distance between a pair of system states increases during the dynamics, indicating an increase in their distinguishability and corresponding to an information backflow from the environment.  The maximization is taken over all possible pairs of initial states. However, it has been shown that for qubit states, as is the focus of this work, the states that maximise Eq.~\eqref{eq:continous BLP} are always pure, orthogonal states~\cite{OptimalNM_states}. For a collision model which simulates the open system dynamics via discrete time steps, we shall use a discretized version of Eq.~\eqref{eq:continous BLP} such that the non-Markovianity is quantified as~\cite{ruariMauro, BLPmeasure2},
\begin{equation}
\label{eq:NMdiscrete}
    N = \max_{\{\rho_{0}, \pi_{0}\}} \sum_{n} \left[D(\rho_{n},\pi_{n})-D(\rho_{n-1},\pi_{n-1}) \right].
\end{equation}

\subsection{Steady State Properties for Non-Markovian Collision Models}
\begin{figure}[t]
    \includegraphics[width=0.99\linewidth]{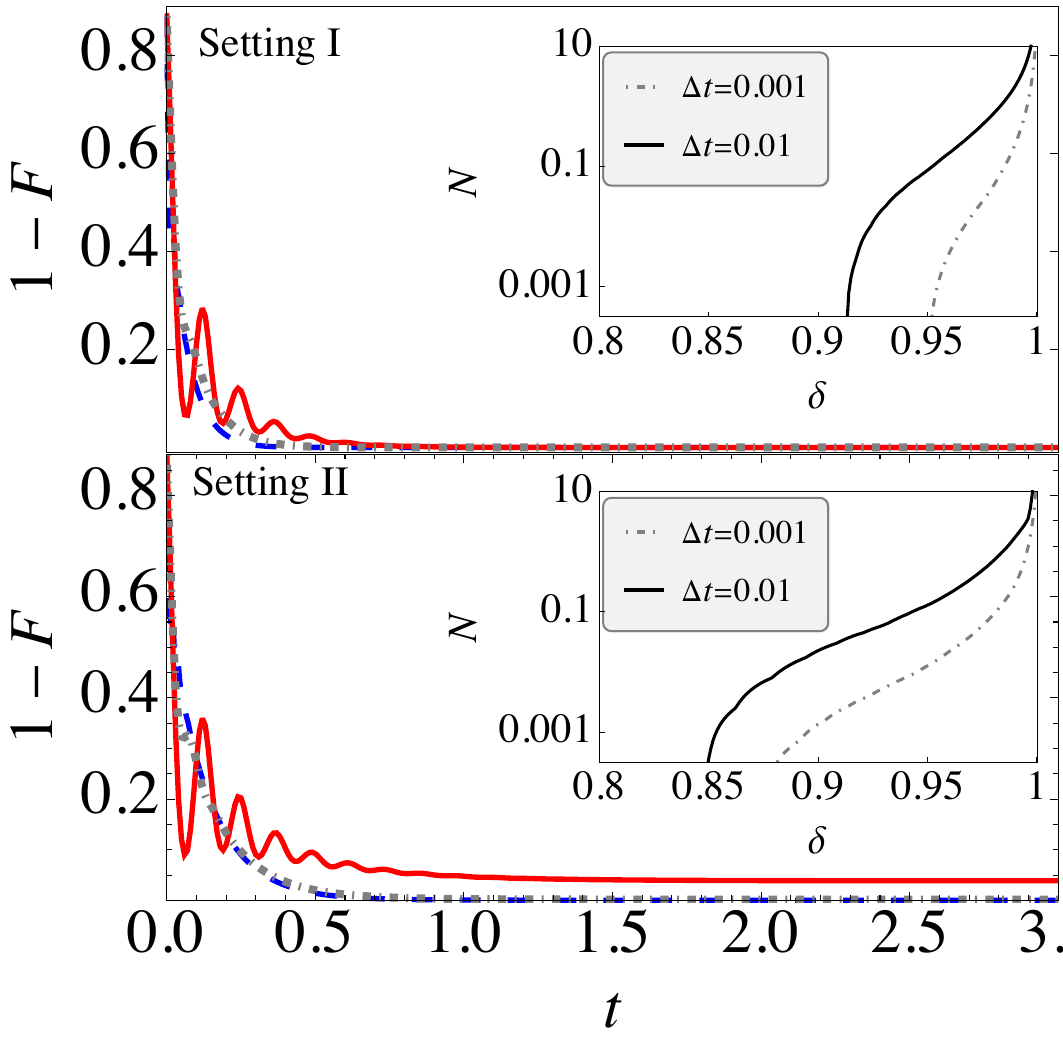}
    \caption{Dynamical approach to the steady state. Upper panel is for setting I and the lower panel is for setting II. We show the fidelity, ($1-F$), against $t \!=\! n \Delta t$, between the evolved system state and Eq.~\eqref{eq:Gibbsstate} fixing $\beta\!=\!2$. In both settings we consider three sets of parameters: Solid, red lines correspond to long system-environment collision durations, $\Delta t\!=\!0.01$, and strong intra-environment interactions $\delta\!=\!0.95\tfrac{\pi}{2}$; Dashed, blue curves correspond to long system-environment collision durations, $\Delta t\!=\!0.01$, and weaker intra-environment interactions, $\delta\!=\!0.8\tfrac{\pi}{2}$; Gray, dot-dashed curves correspond to short system-environment collision durations $\Delta t\!=\!0.001$ and strong intra-environment interactions, $\delta\!=\!0.95\tfrac{\pi}{2}$. The insets show the discretized non-Markovianity measure, Eq.~\eqref{eq:NMdiscrete}, evaluated for $\beta\!=\!2$ as a function of intra-environment interactions strength $\delta$. In both main panels the system is initialized in the excited state $\rho_0 = \ketbra{1}{1}$. Other parameters used: $\omega =1, \Gamma =4.$}  
    \label{fig:NMFidelityLinePlots}
\end{figure}

We begin by considering how the system approaches the steady state in each setting. For concreteness, we set $\beta\!=\!2$ which fixes initial auxiliary unit states for setting I and the interaction strengths, given in Eq.~\eqref{eq:TwoBathInts}, for setting II. The most relevant parameters dictating the dynamics are then the $S$-$A_n$ interaction time $\Delta t$ and the strength of the intra-environment collisions, $\delta$. In Fig.~\ref{fig:NMFidelityLinePlots} we examine the impact that these parameters have on the ensuing dynamics and the distance from the expected thermal state. We show the fidelity between the evolved state and Eq.~\eqref{eq:Gibbsstate},
\begin{equation}
\label{eq:Fidelity}
F = \left(\text{Tr}\left[\sqrt{\sqrt{\rho^\text{th}}\rho_n \sqrt{\rho^\text{th}}}\right]\right)^2.
\end{equation}

Considering first  $\delta\!=\!0.8\tfrac{\pi}{2}$ and $\Delta t\!=\!0.01$ for setting I (blue dashed line) we see that the system monotonically approaches the expected Gibbs state at temperature $\beta=2$. The dynamics appears effectively Markovian, despite the presence of the additional intra-environment interactions. Increasing the intra-environment interactions close to the maximal value is necessary before clear signatures of non-Markovianity emerge, as captured by the red solid curves in Fig.~\ref{fig:NMFidelityLinePlots} where $\delta\!=0.95\tfrac{\pi}{2}\!$. However, these signatures depend on the value of $\Delta t$. Taking a smaller interaction time we find that, even in the presence of such extremely strong intra-environment interactions, the system's dynamics still appears Markovian (gray dot-dashed line). Indeed, this can be confirmed by explicitly calculating the BLP measure, Eq.~\eqref{eq:NMdiscrete}, shown in the inset. We see that for a given value of $\Delta t$, there is a critical value of $\delta$ below which the BLP measure is zero, and hence the dynamics appears Markovian. These dynamical differences notwithstanding, for all parameter choices in setting I the system is invariably driven to the expected thermal state. This can be readily confirmed by performing the Markovian embedding and calculating the steady state of system and memory~\cite{Bassano}. The nature of the system-bath interactions ultimately leads to homogenisation and, as will be relevant in later discussions, an asymptotic product state between system and bath units~\cite{GiovannettiPRA2018}. 

Turning our attention to setting II, for the same parameters we find some notably different behaviors shown in the lower panel of Fig.~\ref{fig:NMFidelityLinePlots}. For $\delta\!=\!0.8\tfrac{\pi}{2}$ and $\Delta t\!=\!0.01$ (blue dashed line), the approach to the steady state is still monotonic, while for $\delta\!=\!0.95\tfrac{\pi}{2}$ (red line) the fidelity shows sizeable oscillations before settling to the steady state value. Thus, inline with setting I, strong intra-environment interactions are required to clearly signal the non-Markovian nature of the dynamics. We again find that there is a minimum value of $\delta$ below which the non-Markovian nature of the dynamics is not captured by revivals in the trace distance. We remark that while the BLP measure indicates that setting II is ``more'' non-Markovian, in the sense that  smaller intra-environment collision strengths are necessary for revivals to be present, the overall trend is similar for the two settings and the amount of non-Markovianity as quantified by Eq.~\eqref{eq:NMdiscrete} is comparable. More interestingly, however, we now see that in the presence of such strong intra-environment interactions the system's steady state is substantially different from the expected thermal state, as captured by the solid red line in the lower panel of Fig.~\ref{fig:NMFidelityLinePlots} settling to a non-zero value.

It follows then that for sufficiently weak intra-environment interactions, aligning with parameter ranges where suitable measures of non-Markovianity fail to reveal any strong memory effects, settings I and II appear essentially equivalent at the level of the steady states insofar as that they yield the same thermal steady state. Setting I, by construction, will drive the system to the Gibbs state at inverse temperature $\beta$ for any value of $\Delta t>0$ and $0\!\leq\! \delta\! <\!\tfrac{\pi}{2}$~\cite{GiovannettiPRA2018}.
By contrast, in the presence of strong intra-environment interactions setting II drives the system to a NESS which is in general at a different temperature. The deviation of the NESS in setting II from the canonical thermal state becomes more significant as the degree of non-Markovianity is increased.

\begin{figure}[t] 
    \centering
    (a) \hskip0.5\linewidth (b)\\
    \includegraphics[width=0.49\linewidth]{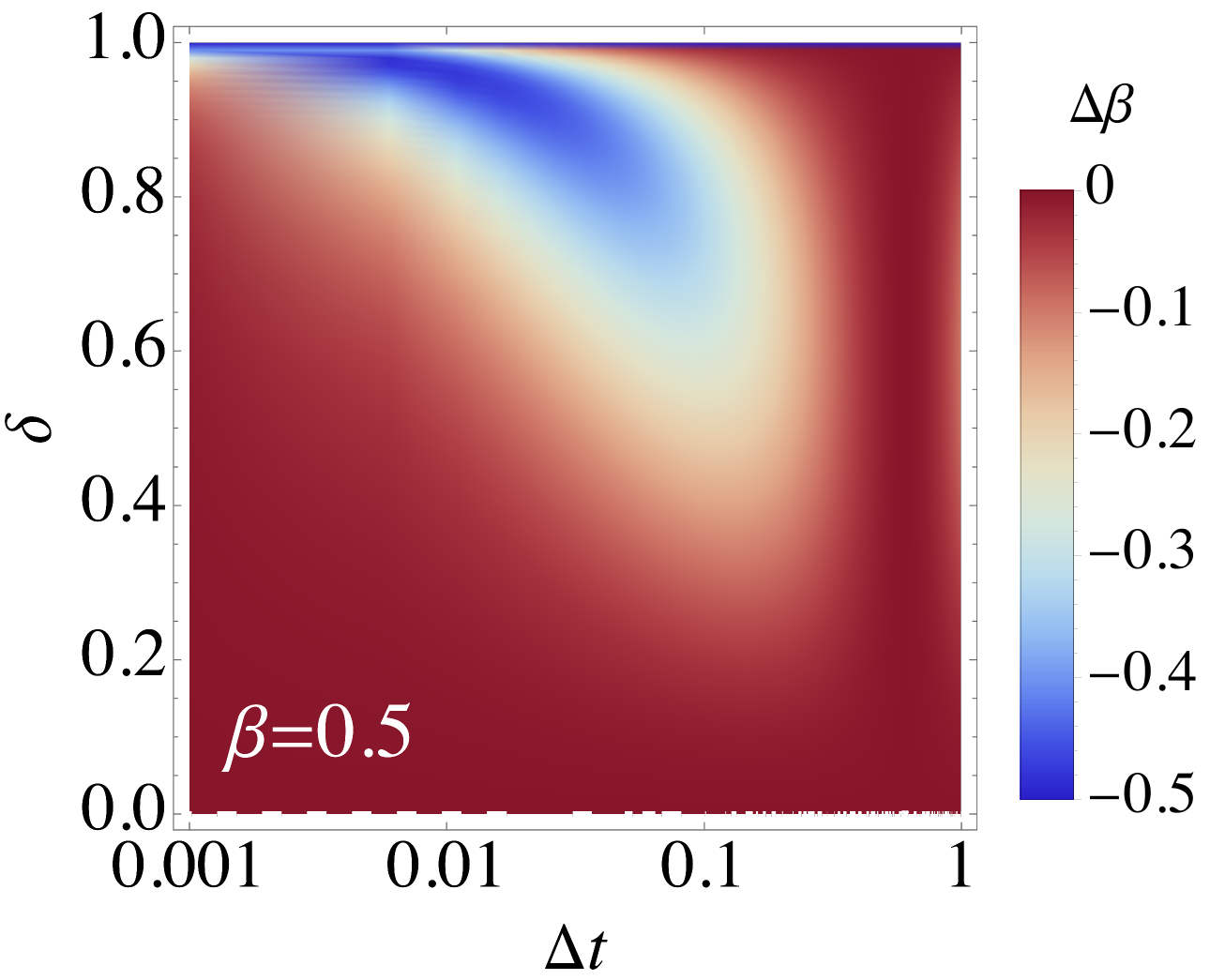}
    \includegraphics[width=0.49\linewidth]{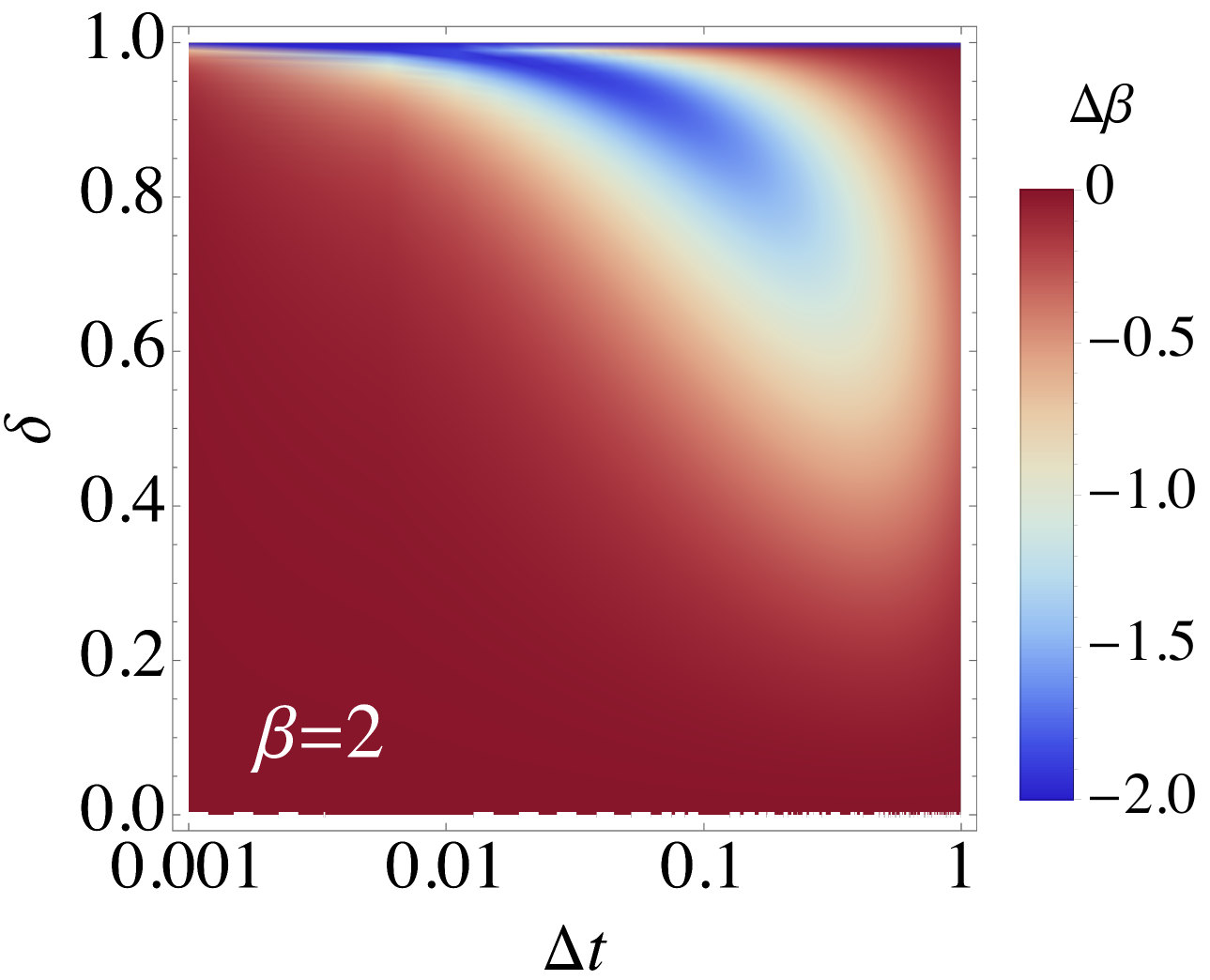}\\
    (c) \hskip0.5\linewidth (d)\\
    \includegraphics[width=0.49\linewidth]{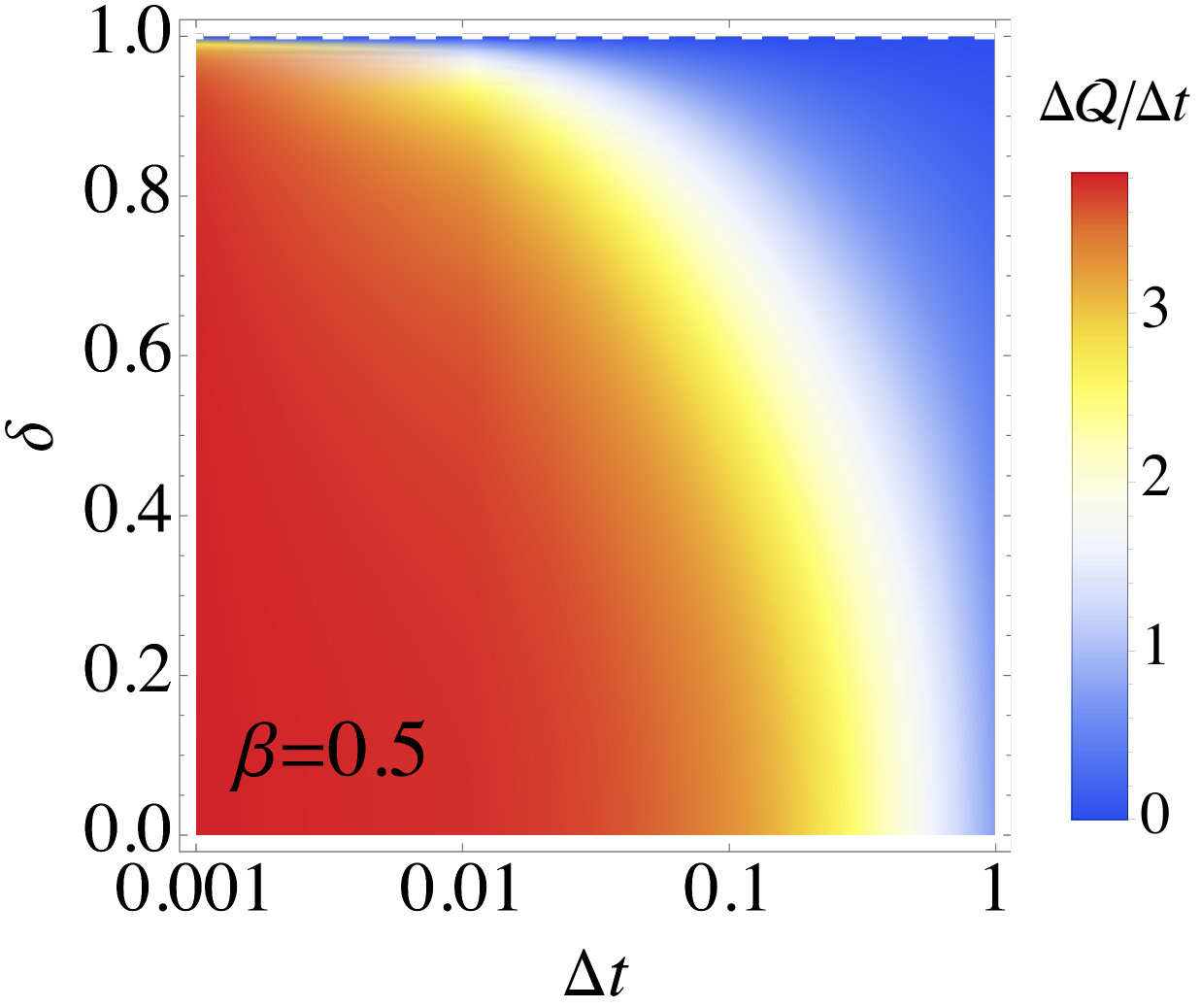}
    \includegraphics[width=0.49\linewidth]{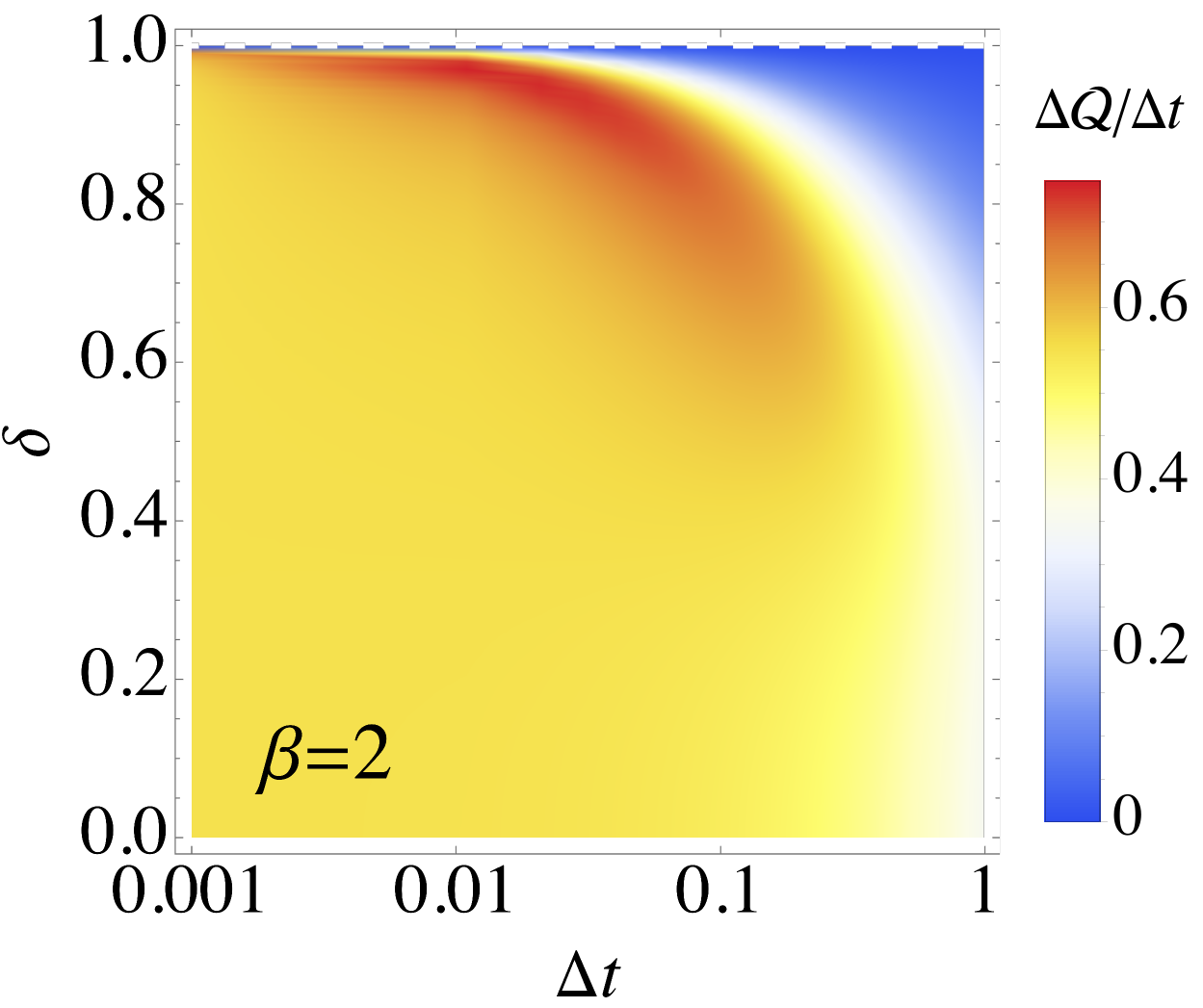}
    \caption{Panels (a) and (b): Difference in system temperature reached at the steady state, $\Delta\beta\!=\!{\beta_{e}-\beta}$ considering the impact of increasing the A-A interactions $\delta$ and increasing the S-A collision duration $\Delta t$. Panels (c) and (d): Associated steady state heat flux, $\Delta Q / \Delta t$, into $A_n^{(0)}$. Other parameters used: $\omega\!=\!1, \Gamma\!=\!4$.}
    \label{fig:FidelityDensityPlots}
\end{figure}
In Fig.~\ref{fig:FidelityDensityPlots} we consider the impact of varying $\delta$ and $\Delta t$ on the resulting steady state in setting II, where we consider two values of temperature, $\beta\!=\!0.5$ and $2$. As the steady state is diagonal, we can associate to it an effective temperature, $\beta_e$. This is formally defined as follows: as the state is diagonal, it can always be written in the form~\eqref{eq:Gibbsstate}, with an effective parameter $g_e$. The effective temperature is then defined as 
\begin{equation}
\label{eq:efftemp}
    \beta_e = \frac{1}{\omega}\log \left( \frac{1+g_e}{1-g_e}\right).
\end{equation} 
In panels (a) and (b) we show $\Delta \beta=\beta_e-\beta$, i.e.  the difference between the effective temperature $\beta_e$ that the steady state reaches and the canonical temperature $\beta$ appearing implicitly in Eq.~\eqref{eq:Gibbsstate}. We see a non-trivial interplay between the impact that the two parameters have on the resulting steady-state. For example, stronger non-Markovianity $\delta$ is required to see significant deviations from the expected thermal state when the collision time $\Delta t$ is small.
We see that the regions of largest deviation correspond to values of $\Delta\beta\!<\! 0$, i.e.~the steady state is at a {\it higher} temperature than expected. Considering a larger value for $\beta$ itself, panel (b), we see a shrinking of the regions of most substantial deviation. These areas are generally pushed to higher values of $\delta$ and $\Delta t$, indicating that the non-equilibrium character of the steady state vanishes in the zero temperature limit. This is consistent with the fact that for $\beta\!\to\!\infty$, setting II reduces to the single bath case. 

The energy-preserving nature of the interactions means that essentially the same heat current is observed due to all the collisions, i.e.~the heat flux into the cooler bath is equal and opposite to the heat flux out of the excited state bath and, furthermore, the same magnitude of flux into and out of the auxiliary units is observed due to the intra-environment collisions. As a representative example, we show the heat flux into $A_n^{(0)}$ due to the system-bath interaction in Fig.~\ref{fig:FidelityDensityPlots}(c) and (d). As we will see in the following section, these currents aide in the establishment of strong correlations that contribute to explaining why the system settles to a given effective temperature.

\subsection{Steady State Correlations}
We now consider the behavior of system-environment correlations. To do this we must employ a suitable measure and, for our purposes, we choose the negativity based on the postive partial transpose criterion. For a two qubit system, it is defined~\cite{BiPartite_Neg},
\begin{equation}
\label{eq:BipartiteNeg}
    \mathcal{N}_2  = -2 \max\left[0,\lambda_{neg}\right],
\end{equation}
where $\lambda_{neg}$ is the negative eigenvalue of the partially transposed density matrix. As we use  the Markovian embedding to determine the steady state from the non-Markovian dynamics of setting II, our state consists of a three-qubit system made up of $S$ and the two ``memory'' qubits. We can extend Eq.~\eqref{eq:BipartiteNeg} to quantify entanglement between different bi-partitions of the tripartite system. For three qubits $A,B$ and $C$ it follows that the entanglement between $A$ and the combined state of $B$ and $C$ is given by
\begin{equation}
\mathcal{N}_{A(BC)}=-2 \max\left[0,\sum_j \lambda_{neg,j}^{A(BC)}\right],
\end{equation}
where $\lambda_{neg,j}^{A(BC)}$ are the negative eigenvalues of the partially transposed density matrix of the bipartition $A$-$BC$. The genuine tripartite entanglement of the three-qubit state can then be calculated by taking a geometric average of the three possible bipartitions~\cite{sabin2008classification},
$$\mathcal{N}_3 =\left[\mathcal{N}_{A(BC)}\mathcal{N}_{B(AC)}\mathcal{N}_{C(AB)}\right]^{\frac{1}{3}} $$
which provides an entanglement monotone. We remark that this corresponds to a sufficient, but not necessary, condition for entanglement. 

\begin{figure}[t]
    \centering
    (a) \hskip0.5\linewidth (b)\\
    \includegraphics[width=0.49\linewidth]{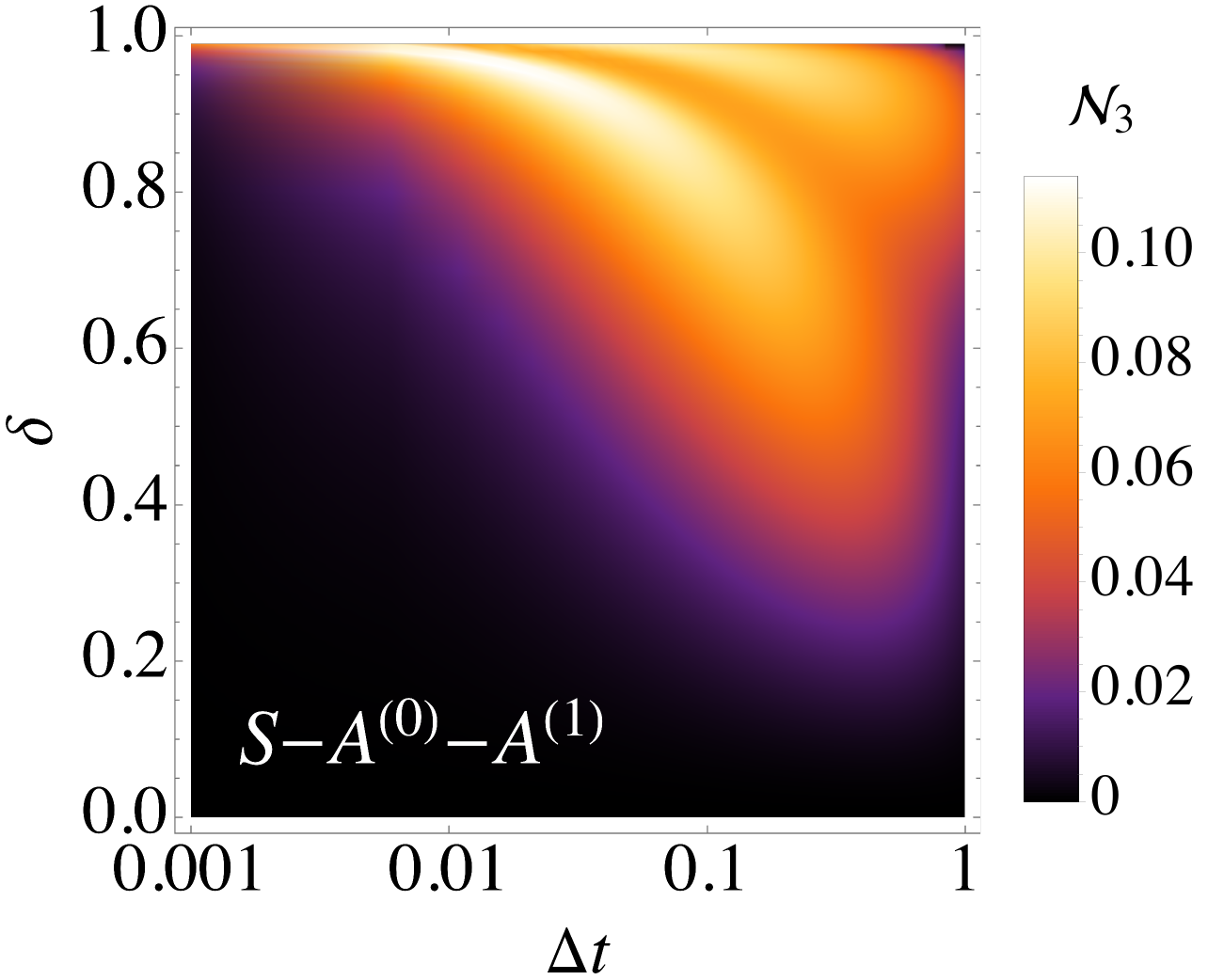}
    \includegraphics[width=0.49\linewidth]{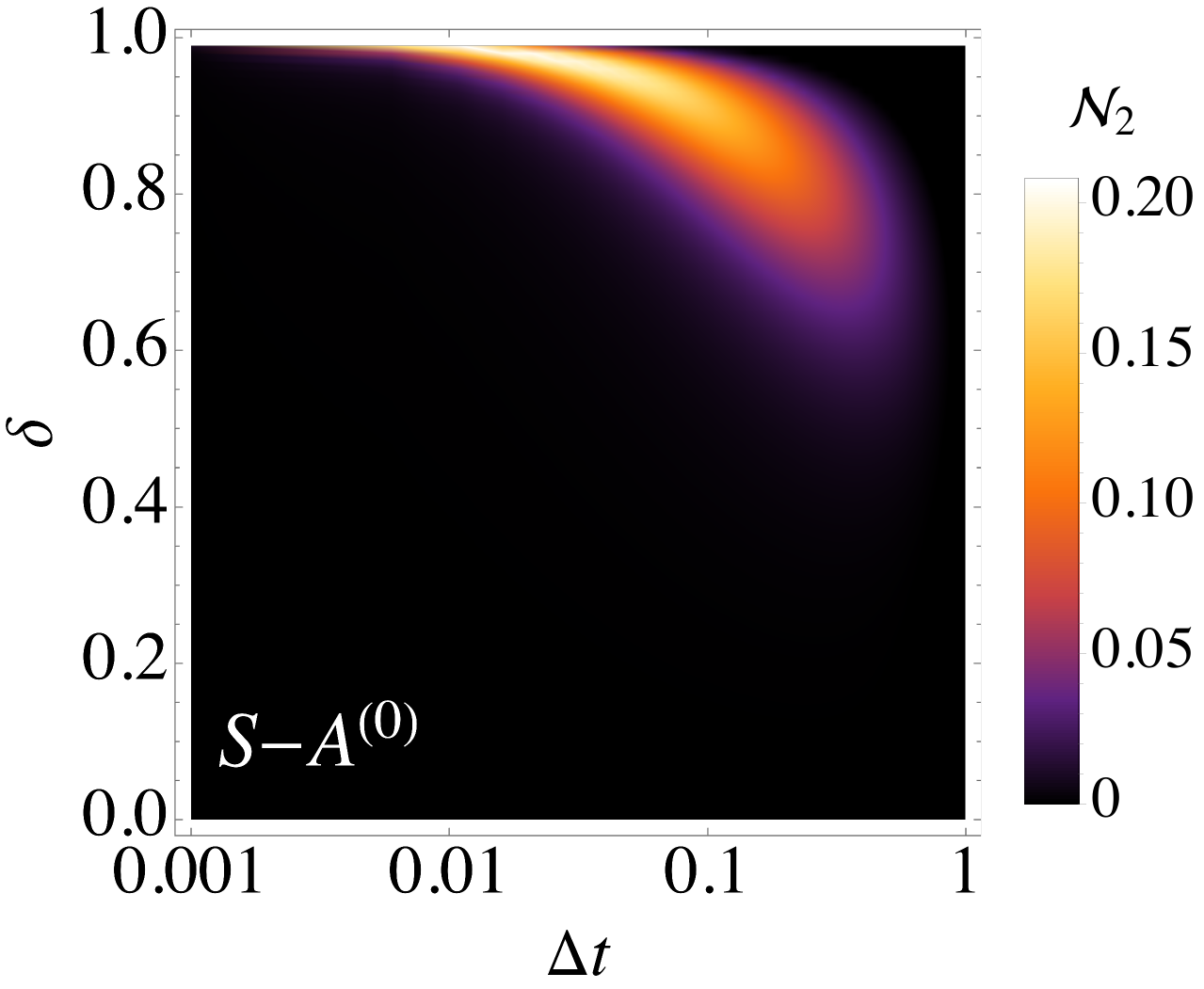}\\
    (c) \hskip0.5\linewidth (d)\\
    \includegraphics[width=0.49\linewidth]{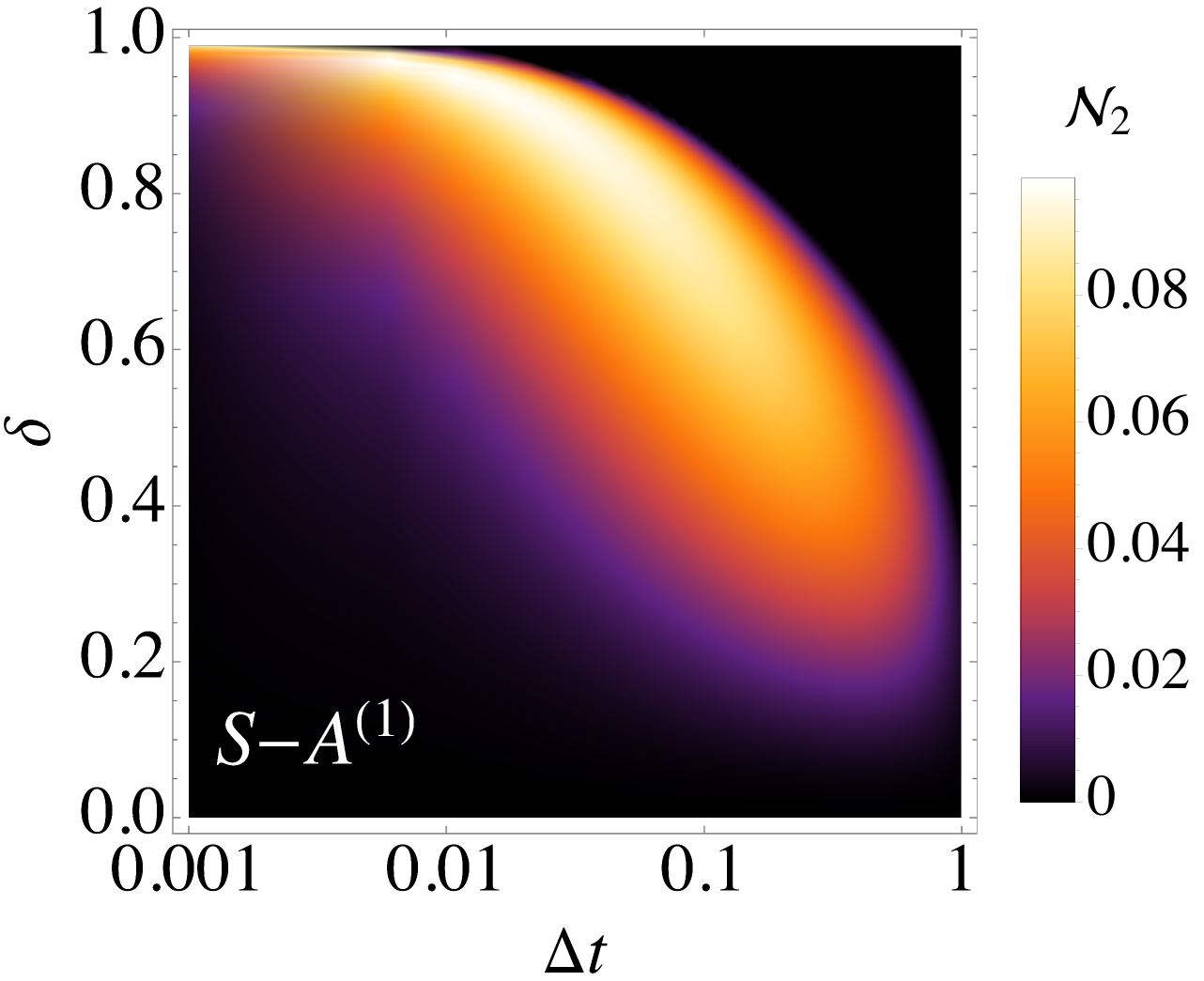}
    \includegraphics[width=0.49\linewidth]{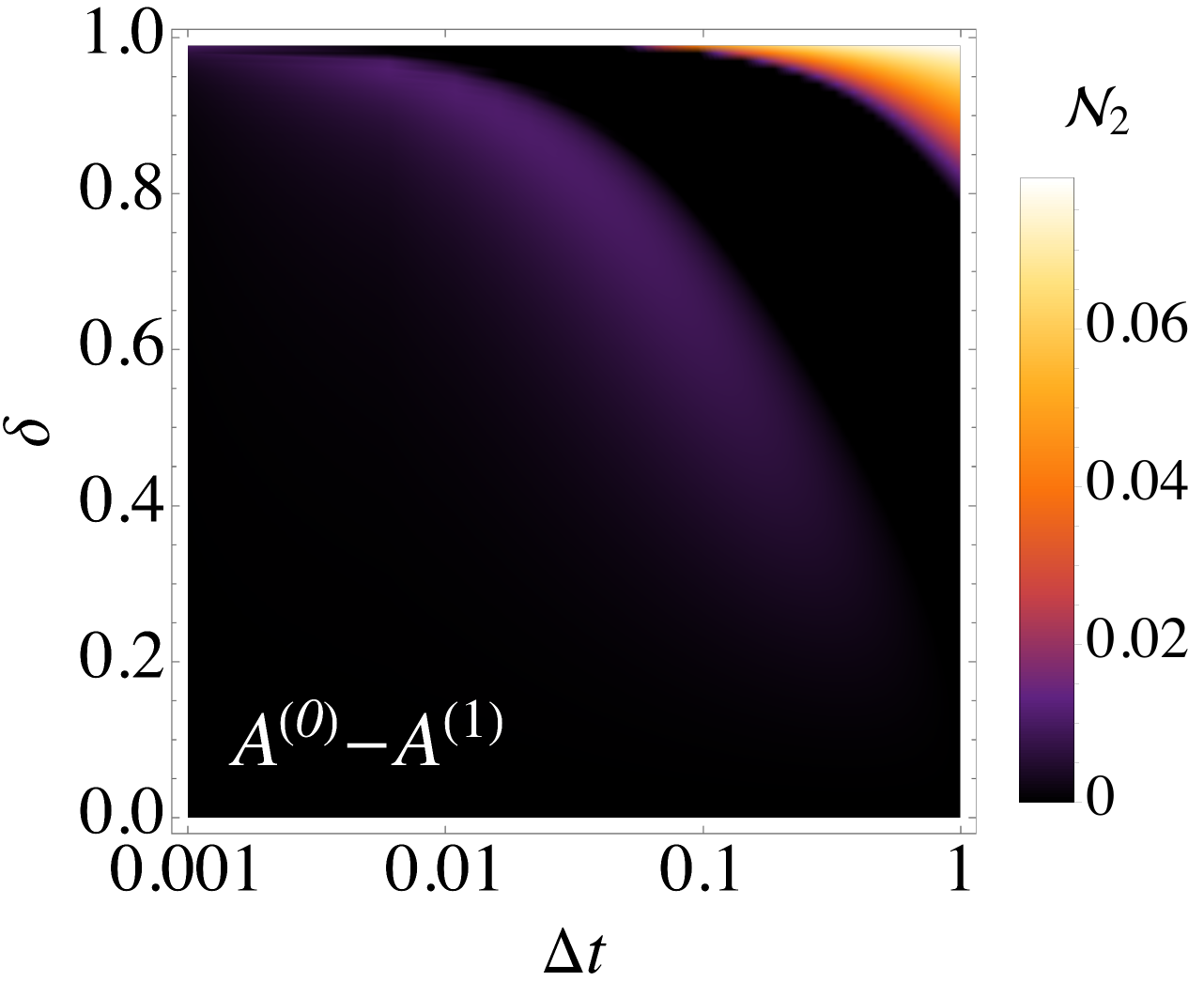}
    \caption{Steady state entanglement arising from setting II. Panel (a): Genuine tripartite entanglement between the system and two ``memory'' units. Panels (b-d): Bipartite entanglement of the various reduced states. In all panels we fix $\beta\!=\!2, \omega =1, \Gamma=4$.}
    \label{fig: Tripartite Negativity}
\end{figure}

In Fig.~\ref{fig: Tripartite Negativity} we examine the entanglement established in the steady state of setting II in the full tripartite state as well as the various reduced state bipartitions. In all cases we find that there is a vanishingly small degree of entanglement present for sufficiently fast system-environment collisions and weak intra-environment interactions, i.e. small $\Delta t$ and $\delta$. Increasing these parameters allows for the establishment of entanglement in the tripartite system as shown in panel (a). The region of non-zero entanglement closely aligns with regions where setting II's steady state significantly deviates from the expected thermal state. Considering the reduced state of the system and one of the baths, shown in panels (b) and (c), we again see a good agreement between regions where the system's steady state is at a different effective temperature and the presence of bipartite entanglement between $S$ and $A_n^{(i)}$. For the reduced state between the two bath units, shown in panel (d), we see that entanglement can be established, however, it becomes appreciable only for relatively large values of both $\Delta t$ and $\delta$. 

It is worth reminding that the above is in stark contrast to setting I's steady state, which is driven to a factorised state between $S$ and $A_n$ and therefore clearly there is no entanglement present in the steady state~\cite{GiovannettiPRA2018}. This can be intuitively understood due to the fact that for the considered interactions, setting I leads to homogenization and therefore in the steady state the system and all bath auxiliary units are all in identical states and, as the intra-environment collisions correspond to a partial swap, they do not change the states. As demonstrated by Fig.~\ref{fig: Tripartite Negativity}, this is no longer the case in setting II, where the system evidently cannot homogenize and therefore each collisions is, generally, a strongly entangling operation. 

We note that the entanglement in the tripartite NESS of system and memories is distinct from entanglement that is created during each collision of the memories with the incoming bath auxiliaries. Indeed, even in the Markovian case, once the system has reached the steady state, each subsequent collision in setting II will lead to the creation of genuine tripartite entanglement between the system and the bath units while leaving the reduced state of the system unchanged. This implies that the deviation from the expected thermal state in the non-Markovian case is a consequence of setting II's ability to create entanglement in the system memory compound in addition to the entanglement being continually created and destroyed in each subsequent collision with the baths. We therefore conclude that the presence of non-Markovianity enhances the non-equilibrium character of the state due to the map's ability to create entanglement. However, we remark that as setting I has a comparable degree of non-Markovianity present in the dynamics (see Fig.~\ref{fig:NMFidelityLinePlots}), it follows that its presence is not solely responsible for the observed deviation from the expected thermal state and rather it is a more complex interplay between the constant energy exchanges, the creation and maintenance of strong quantum correlations, and the non-Markovian character of the dynamics.


\section{Stochastic Heat Currents from Trajectories}
\begin{figure}[t]
    \centering
    \includegraphics[width=0.99\linewidth]{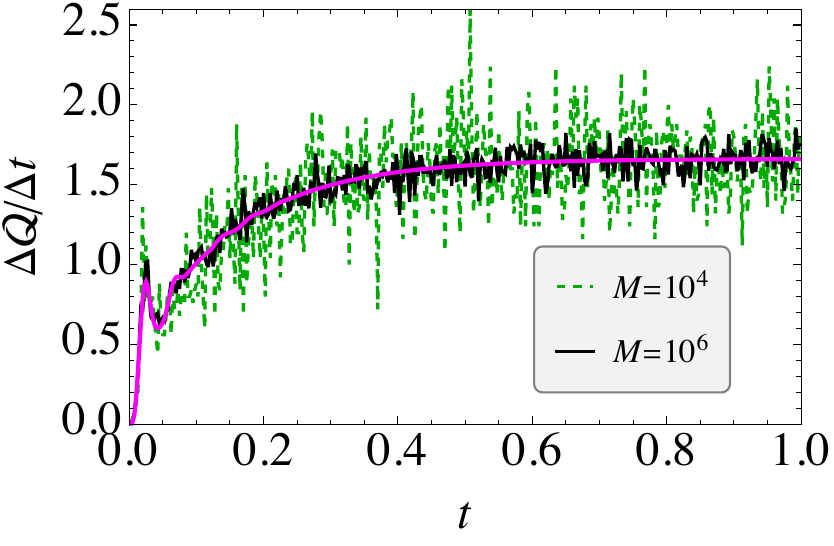}
    \caption{Mean stochastic heat current into the zero temperature bath, $\langle\Delta\tilde{\mathcal{Q}}_{n}^{(0)}\rangle_M$  against $t \!=\! n \Delta t$ for $M=10,000$ (green, dashed line) and $M=100,000$ (black line) along with the corresponding unconditional heat current $\Delta \mathcal{Q}^{(0)}_n$ (magenta line). In all cases the system is initialized in the ground state $\rho_0 = \ketbra{0}{0}$. Parameters used: $\delta = 0.95 \frac{\pi}{2}, \Gamma=4,\beta=1$, $\Delta t=0.0025$.}
    \label{fig:StochasticHeat}
\end{figure}

Quantum collision models lend themselves naturally to the simulation of quantum trajectories when we consider performing measurements on the auxiliary units immediately after their interaction with the system~\cite{Brun_CM_Trajectories}. It is known that even for non-Markovian collision models, individual trajectories can be averaged to recover the unconditional dynamics \cite{Kretschmer_CompositeCM,Whalen_NMCM_Trajectories, cilluffo2021microscopic}. In this section we will show how the unconditional heat current can be recovered from two-point measurements on the auxiliary qubits. The stochastic heat current after one complete collision is calculated by performing a projective measurement, with measurement operator $P_j$, in the energy eigen-basis of the auxiliary unit. This measurement is performed twice, first before the auxilliary unit has undergone any interaction with any other system, and  again after it has completed all of its interactions. This procedure is commonly referred to as a two-point measurement (TPM) scheme. The post-measurement states of the auxiliary unit after the first and second measurements are given by,
\begin{equation}
    \rho_{A^{(i)}_{n}}' = \frac{P_j \rho_{A_n^{(i)}}P_j}{\text{Tr}\left[P_j \rho_{A_n^{(i)}}\right]}, \quad \rho_{A^{(. i)}_{n}}'' = \frac{P_j \tilde{\rho}_{A_n^{(i)}}'P_j}{\text{Tr}\left[P_j \tilde{\rho}_{A_n^{(i)}}'\right]},
\end{equation}
where $\tilde{\rho}'_{A_{n}^{(i)}}$ is the state of the auxiliary unit after it has undergone the first measurement and been involved in the the $A^{(i)}_{n-1}$-$A^{(i)}_{n}$, $S$-$A^{(i)}_n$ and $A^{(i)}_n$-$A^{(i)}_{n+1}$ collisions. The stochastic heat current for the $n$-th time step is then
\begin{equation}
    \Delta\mathcal{Q}_\text{stoc}^{(i)}(n) = \text{Tr}\left[H_{A^{(i)}_n}\left(\rho''_{A_{n}^{(i)}} - \rho_{A^{(i)}_{n}}'\right)\right].
\end{equation}
The unconditional heat current can be recovered from this stochastic current by averaging over many trajectories at each step $n$,
\begin{equation}
    \Delta\mathcal{Q}^{(i)}(n) \simeq\langle\Delta\mathcal{Q}^{(i)}(n)\rangle_M = \frac{1}{M} \sum_{k=1}^M   \Delta\mathcal{Q}^{(i)}_\text{stoc}(n,k),
\end{equation}
where the averaging is done over $M$ trajectories. To give an illustrative example, in Fig.~\ref{fig:StochasticHeat} we calculate the stochastic heat current for the zero temperature bath in setting II for two different values of $M$ and compare it to the unconditional heat current calculated using Eq.~\eqref{eq:Heat}. It should be noted that since in this setting the zero temperature bath starts in the $\ket{0}$ state, the first measurement in the TPM scheme has no effect and the TPM measurement scheme is equivalent to a typical continuous monitoring in this set-up. We see close agreement between the averaged stochastic heat and the unconditional value, with this agreement improving with increasing $M$.

Finally, it is interesting to note that certain heat fluxes in the non-Markovian setting, specifically those associated with the $S$-$A_n$ collisions, cannot be accessed using this type of TPM scheme. For example, in order to access the heat flux specific to the $S$-$A_n^{(0)}$ interaction a measurement must be performed on $A_n^{(0)}$ {\it after} the $S$-$A_n$ interaction, but {\it before} the $A_n^{(0)}$-$A_{n+1}^{(0)}$ interaction takes place. While such a measurement can be done, clearly it will affect the correlations shared between the $S$ and $A_{n+1}^{(0)}$ as well as the incoming state of $A_{n+1}^{(0)}$. This leads to a significant change in system dynamics and thus we will have a different stochastic heat current compared to the unconditional dynamics. These ``inaccessible currents'' appear to be a unique feature of the non-Markovian setting.

\section{Conclusions}
We have examined the steady state properties of a two-level quantum system undergoing an open dynamics simulated using quantum collision models, critically assessing the differences between models that employ a single thermal bath and multiple pure-state baths. As a benchmark, we considered the Markovian case of a simple thermalisation process which can be effectively modelled via a standard GKSL master equation and we leverage the flexibility of collision models to extend this to the non-Markovian setting. We have also shown that the two-bath setup allows for non-Markovian quantum trajectories to be defined. The two-bath setup therefore combines the flexibility of collision models to introduce nontrivial memory effects, with the ability to resolve interpretable stochastic quantities via TPM schemes. Our work builds on recent studies that have considered the steady state properties of Markovian collision models~\cite{Guarnieri2020PLA, Segal2025QST, prositto2025collisional} and extends them to the multi-bath and non-Markovian regime

We considered the typical situation, referred to as setting I, where the collisional bath consists of identical auxiliary units all prepared in the same thermal state. For the system-environment interactions given by Eq.~\eqref{eqn: interaction H}, setting I is known to lead to thermalisation, leaving the system in equilibrium with the bath and, furthermore, the system and environment share no correlations in the steady state. By contrast, setting II involves two collisional baths, whose auxiliary qubits are prepared in either their ground or excited states, and the bath temperature is encoded through the system-bath couplings.  
As a main result we have demonstrated that despite both settings recovering the GKSL master equation for the system in the $\Delta t\!\to\! 0$ limit, setting II necessarily drives the system to a nonequilibrium steady state (NESS), supported by non-zero steady state heat currents. By exploiting the flexibility of collision models, we examined the impact  of non-Markovian dynamics on the resulting steady state, by introducing additional intra-environment interactions. Although setting I and setting II exhibit similar amounts of non-Markovianity, as quantified via the information back flow, we showed that the non-Markovian case of setting II results in the system being driven away from the expected thermal state. Although setting I always reaches the canonical temperature of the bath, the non-equilibrium steady state of setting II can reach a different temperature. Through a careful analysis of the tripartite and bipartite entanglement, we have established that the non-equilibrium character of the steady state is rooted in the ability of setting II to create and maintain strong quantum correlations. We have also demonstrated how the average heat current can be recovered in setting II using a two-point measurement scheme, even in the presence of non-Markovianity. 

Our results provide insight into the dynamics and thermodynamics of thermalisation and equilibration and highlight that seemingly equivalent dynamics at the level of the system of interest can correspond to fundamentally different physical processes. The versatility of the collision model framework allows them to be employed in the modeling of practical quantum devices and offers a promising avenue for future work. One possible direction is to model non-Markovian \textit{fermionic} environments relevant to semiconductor quantum dot devices coupled to electronic circuits~\cite{goldhaber1998kondo,shim2018numerical,barthelemy2013quantum,iftikhar2018tunable,pouse2023quantum,petropoulos2024nanoscale}, where we expect memory effects and strong system-bath correlations to play an important role. This builds on recent work showing that non-Markovian collision models are a powerful and systematic method for treating bosonic environments~\cite{Lacroix2024MakingQCMExact}. Another possibility would be to use the framework of non-Markovian collisional trajectories for measurement-based feedback control in non-Markovian environments, for example, to see the effects of specific intra-environment interactions and the resulting environmental memory on the ability to prepare a given target state~\cite{HarwoodCMFeedback}.

\section*{Appendix}

\subsection{Continuous-time limit of setting II}
Following similar steps as Refs.~\cite{CICCARELLO20221, cusumano2022quantum} we can show that GKSL master equation, Eq.~\eqref{Eqn: QME} is recovered in the time-continuous limit also for setting II. We assume an arbitrary form for the system-auxiliary interaction
\begin{equation}
    H_I = \sum_\nu g_\nu A_\nu B_\nu
\end{equation}
where $g_\nu$ are coefficients, $A_\nu$ and $B_\nu$ are sets of system and auxiliary operators, respectively, and $\Delta\rho_n = \rho_n - \rho_{n-1}$. By expanding the unitary operator governing a single collision up to second order in $\Delta t$, it can be readily shown that a finite difference equation for the state of the system can be derived~\cite{CICCARELLO20221}
\begin{multline}
\label{finitediffME}
   \frac{\Delta \rho_n}{\Delta t} = -i[H_S+\sum_\nu g_\nu \langle B_\nu\rangle A_\nu,\rho_{n-1}]\\
   + \sum_{\nu\mu}g_\nu g_\mu \Delta t\langle B_\mu B_\nu\rangle(A_\nu \rho_{n-1}A_\mu-\frac{1}{2}\{A_\mu A_\nu,\rho_{n-1}\}
\end{multline}
To arrive at the master equation for the system it therefore follows that we require the first and second moments of the auxilliary operators 
\begin{equation}
    \langle B_\nu\rangle=\text{Tr}[B_\nu \eta_n], \qquad  \langle B_\nu B_\mu\rangle=\text{Tr}[B_\nu B_\mu \eta_n].
\end{equation}
where we have introduced the state of the $n$-th auxiliary, $\eta_n$. Note that thus far we have made no assumptions about the specific form of interaction or the structure of the auxiliary. For setting II, we assume the auxiliary is $\eta_n\!=\!(\ketbra{0}{0} \otimes \ketbra{1}{1})$ and for convenience we rewrite the interaction as
\begin{equation}
H_I = \sum_{i=0,1} -J^{(i)}(\sigma_S^+ \otimes \sigma_{A^{(i)}_n}^- +\sigma_S^- \otimes \sigma_{A^{(i)}_n}^+).
\end{equation} 
It then follows that we need only compute the relevant moments with,
\begin{align*}
    A_\nu &\in \{\sigma_S^+,\sigma_S^-,\sigma_S^+,\sigma_S^-\} \\
    B_\nu &\in \{\sigma_{A_{n}^{(0)}}^- ,\sigma_{A_{n}^{(0)}}^+,\sigma_{A_{n}^{(1)}}^-,\sigma_{A_{n}^{(1)}}^+\} \\ \
    g_\nu &\in \{-J^{(0)},-J^{(0)},-J^{(1)},-J^{(1)}\}.
\end{align*}
For this set of operators and specific form of $\eta_n$, we find $\sum_\nu g_\nu \langle B_\nu\rangle A_\nu =0$, while the only non-zero second moments are $\langle \sigma_{A_{n}^{(0)}}^- \sigma_{A_{n}^{(0)}}^+ \rangle = \langle \sigma_{A_{n}^{(1)}}^+ \sigma_{A_{n}^{(1)}}^- \rangle =1$. Equation~\eqref{finitediffME} then reduces to,
\begin{multline}
     \frac{\Delta \rho_n}{\Delta t} = -i[H_S,\rho_{n-1}] + (J^{(0)})^2\Delta t\mathcal{D}[\sigma_S^-]\rho_{n-1} \\
     +  (J^{(1)})^2\Delta t\mathcal{D}[\sigma_S^+]\rho_{n-1}.
\end{multline}
In order for the dissipators to survive the continuous time limit we introduce diverging coupling strengths as given in the main text
\begin{equation}
J^{(0)} = \sqrt{\frac{\Gamma(\bar{N}+1)}{\Delta t}}, \quad J^{(1)} = \sqrt{\frac{\Gamma \bar{N}}{\Delta t}},
\end{equation}
such that after taking  $\Delta t \rightarrow 0$ we obtain,
\begin{equation}
    \dfrac{d\rho}{dt} =  -i \left[H_S,\rho \right]+\Gamma(\bar{N}+1)\mathcal{D}[\sigma^-]\rho +\Gamma\bar{N}\mathcal{D}[\sigma^+]\rho \;,
\end{equation}
i.e. the GKSL master equation given in Eq.~\eqref{Eqn: QME}.

\subsection{Heisenberg system-environment interaction}
Here we consider how changing the interaction between system and environment impacts the resulting steady states arising from the two collision model settings. In particular, we consider an isotropic Heisenberg interaction
\begin{equation}\label{eqn: interaction HXXX}
    H_{I} = -\frac{J}{2}\left(\sigma^x_S \otimes \sigma^x_{A_n} + \sigma^y_S \otimes \sigma^y_{A_n} + \sigma^z_S \otimes \sigma^z_{A_n}\right) \;,
\end{equation}
which gives rise to precisely the partial-SWAP operation
\begin{equation}\label{eqn: Markovian Thermal Unitary}
     U_n^{\text{I}} = \cos({J \Delta t})\mathbbm{1} - i\sin({J \Delta t}) \mathcal{W}_{S,A_n}\;,
 \end{equation}
similar to Eq.~\eqref{eq:NMinteraction} in the main text. We remark that, at the level of the steady states, many of the main insights discussed in the main text are unaffected. We therefore focus on only the aspects where differences emerge.

For collision models following setting I, this type of system-environment interaction has been extensively considered~\cite{ScaraniPRL2002, ruariMauro, Bassano} (we remark that one can also consider an incoherent SWAP operation, see for example Ref.~\cite{karpat2025transient}). For setting I, it should be evident that this choice of interaction has no impact on the steady state that the system is driven to; the dynamics will invariably homogenize the system with initial state of the auxiliary units in both the Markovian and non-Markovian regimes. For auxiliaries again initialised in thermal states, i.e. Eq.~\eqref{eq:Gibbsstate}, it known that despite the dynamics driving the system to the same thermal state, for this type of interaction we no longer recover the dynamics governed by Eq.~\eqref{Eqn: QME}~\cite{cusumano2022quantum}. Nevertheless, as our focus is on the steady-state properties our interest lies in examining the how this choice of interaction reveals more substantial differences between settings I and II.

An immediate and remarkable difference emerges even in the Markovian case; i.e. $\delta\!=\!0$. The analytical form of the steady state in setting II is readily accessible, although too cumbersome to provide any direct insight. Nevertheless, it is diagonal in the system's energy eigenbasis and, since we are considering a simple two-level system, it can therefore be assigned an effective temperature, $\beta_{e}$, via Eq.~\eqref{eq:efftemp}. While the explicit formula of $g_e$ or $\beta_e$ is too cumbersome, to illustrate the idea, in the case of very small $\Delta t$ we take a Taylor expansion to first order and find
\begin{equation}\label{no109819uj101u9}
    g_e \simeq g + \frac{\Gamma  \bar{N}(\bar{N}+1)}{6(2\bar{N}+1)^2} \Delta t \;.
\end{equation}
This demonstrates that for system environment interactions governed by the partial-SWAP, even in the Markovian limit setting II drives the system to a steady state that deviates from the equilibrium value $g = (2\bar{N}+1)^{-1}$ by a term that is proportional to $\Delta t$.

\begin{figure}[t]
    (a) \\
\includegraphics[width=0.7\linewidth]{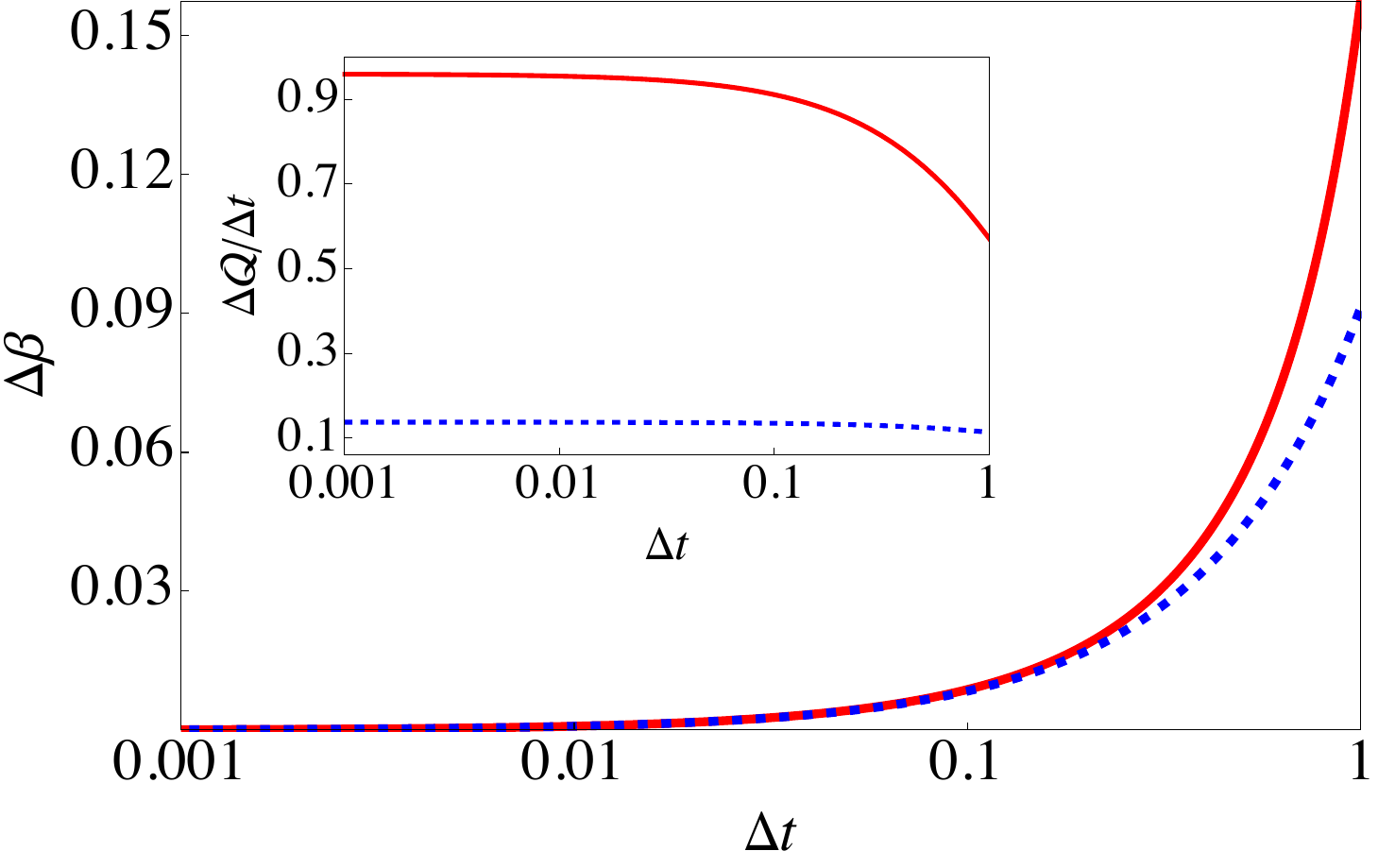}\\
    (b) \hskip0.5\linewidth (c)\\
    \includegraphics[width=0.49\linewidth]{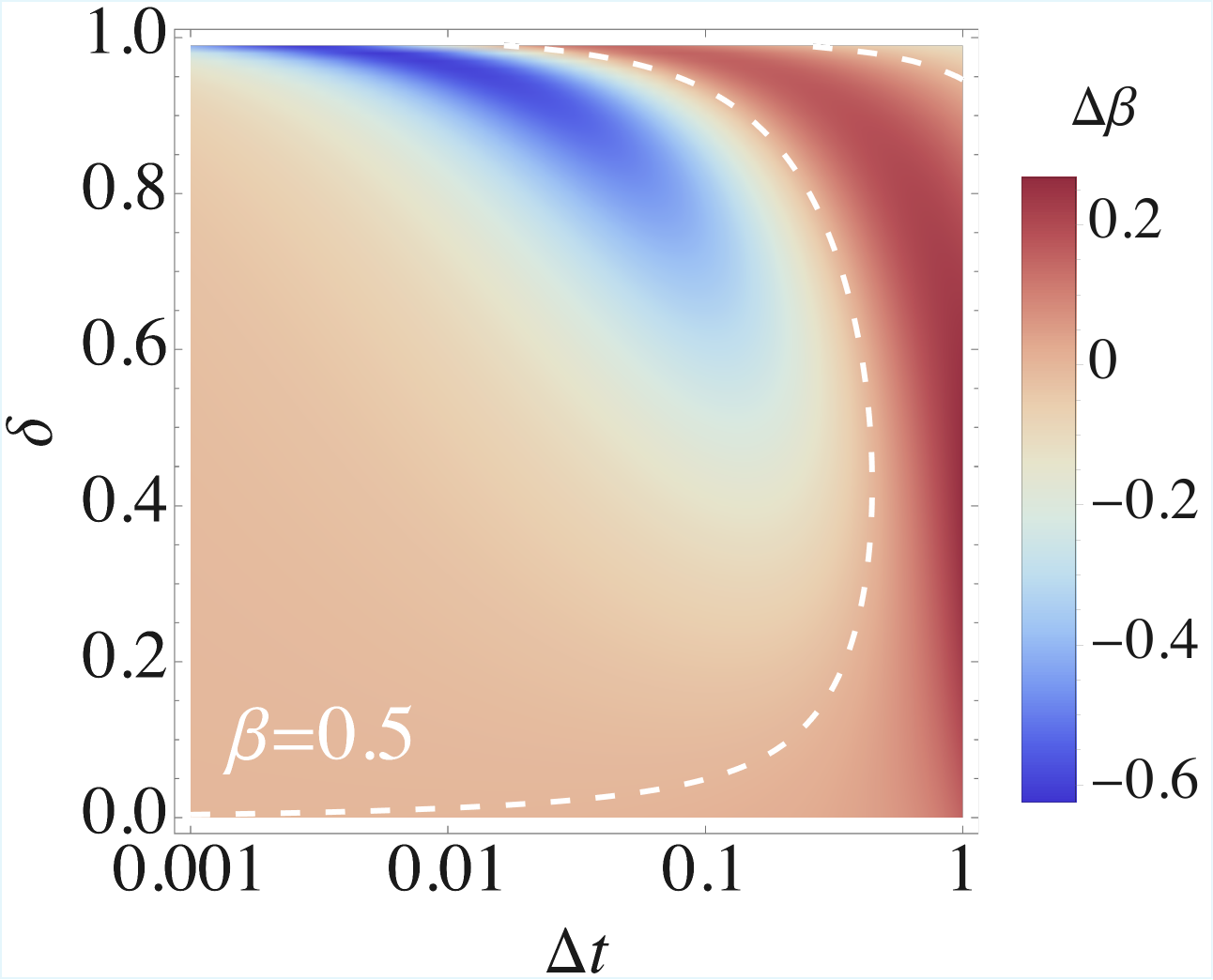}
    \includegraphics[width=0.49\linewidth]{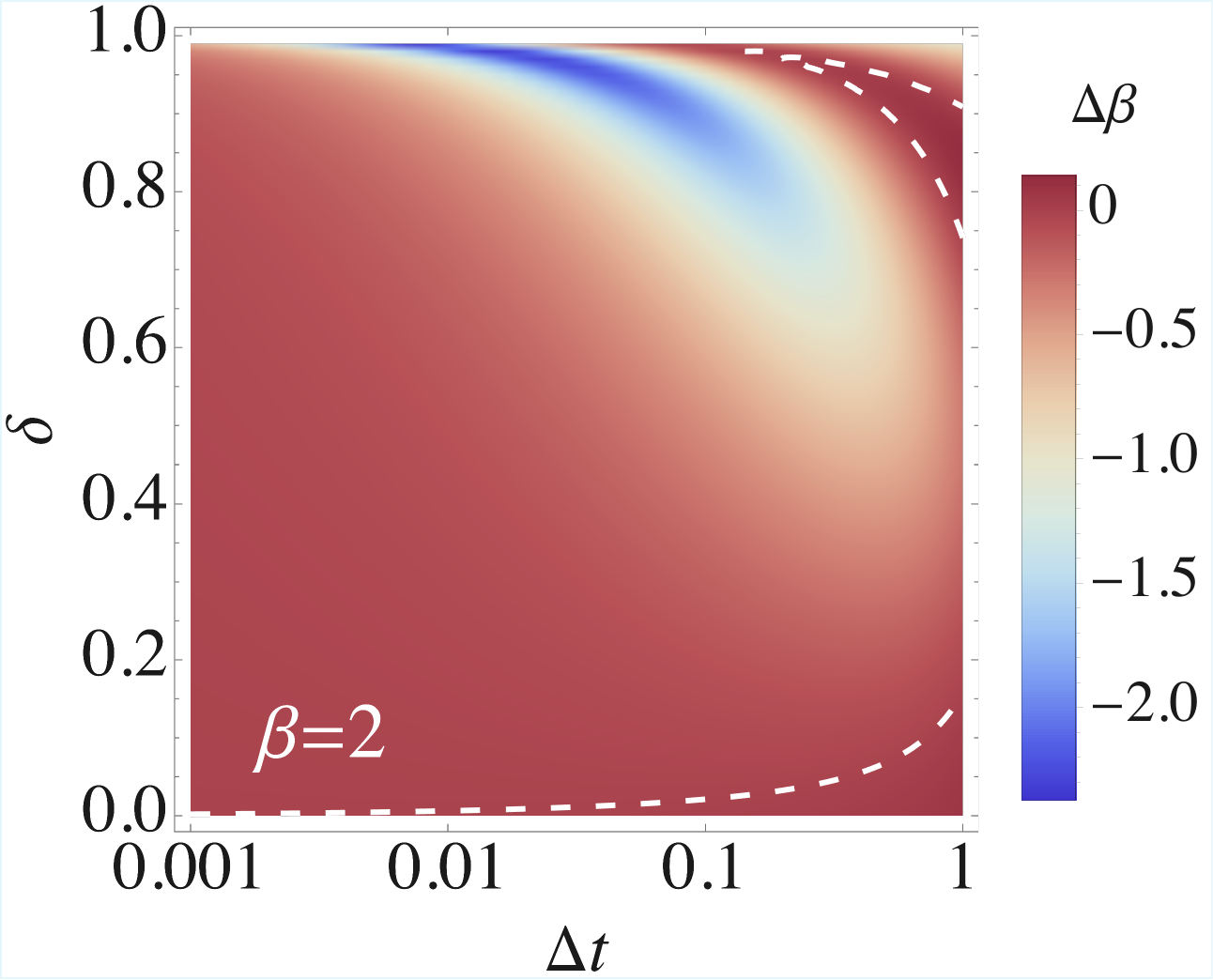}\\
\caption{(a) Steady states in the Markovian limit for system environment interactions given by Eq.~\eqref{eqn: interaction HXXX}. {\it Main panel:} Deviation of the effective temperature from the canonical temperature, $\Delta\beta$, with increasing collision time-step $\Delta t$ for different canonical temperatures $\beta\!=\!2$ (dashed, blue; lower temperature) and $\beta\!=\!0.5$ (solid, red; higher temperature). {\it Inset:} Non-equilibrium steady state heat current, $\Delta\mathcal{Q}/\Delta t$ for the initially low energy (cold) bath, $A_n^{(0)}$. (b) and (c)  Difference in system temperature reached at the steady state, $\Delta\beta\!=\!{\beta_{e}-\beta}$ considering the impact of increasing the A-A interactions $\delta$ and increasing the S-A collision duration $\Delta t$. Other parameters used: $\omega\!=\!1, \Gamma\!=\!1$.}  
\label{fig:Markovian_XXX}
\end{figure}

In Fig.~\ref{fig:Markovian_XXX}(a) we show this deviation from the ``canonical" steady state temperature, $\beta$, showing $\Delta \beta = \beta_{e} - \beta$. We clearly observe that only in the limit of short collisions ($\Delta t \to 0$) is the system is driven to the expected steady state, i.e. $\beta_{e}\!=\!\beta$, as predicted by Eq.~\eqref{no109819uj101u9}. Increasing $\Delta t$, corresponding to longer interaction times during the collisions, leads to a substantial increase in $\Delta \beta$, with this deviation becoming more pronounced for increasing temperatures. Larger $\beta$ leads to larger deviations and we consistently find that $\Delta \beta\!>\!0$, indicating that the effective temperature that the system is driven to is cooler than expected. For completeness, in the inset we show that NESS current for the $\ket{0}$ bath again confirming the non-equilibrium character of the steady state.

In panels (b) and (c) we introduce non-Markovianity via the same intra-environment interactions, Eq.~\eqref{eq:NMinteraction}. While it is clear that the qualitative behavior is consistent with the exchange interaction considered in the main text, we do see some differences worth highlighting. In particular we see that $\Delta \beta$ can be positive or negative depending on interplay between the strength of the intra-environment interactions and the system environment collision duration. This is markedly different from the Markovian case shown in Fig.~\ref{fig:Markovian_XXX}(a) where $\Delta\beta\!>\!0$. This variation in the effective temperature that the steady state reaches, corresponding to a transition from effective heating to cooling, implies that in the non-Markovian two bath setting the system can still reach the expected thermal state, shown by the white dashed lines in Fig.~\ref{fig:Markovian_XXX}(b) and (c) indicating the parameter values where $\Delta\beta\!=\!0$. This provides further evidence that while the non-Markovianity of the dynamics is a contributing factor for the system reaching a non-equilibrium steady state, there is no simple relationship between the degree of non-Markovianity and the effective temperature to which the system settles.

As a final remark, we note that in the Markovian case $\delta=0$, Eq.~\eqref{no109819uj101u9} shows that in the limit $\Delta t \to 0$ the steady state of setting II is the canonical Gibbs state with $g_e\to g$ and hence $\beta_e\to \beta$. Interestingly, in the non-Markovian case, a different effective temperature $\beta_e \ne \beta$ can be reached even in the limit $\Delta t \to 0$. As can be appreciated visually from Fig.~\ref{fig:Markovian_XXX}(b) and (c), the blue regions of large temperature deviation are pushed to larger and larger $\delta$ as $\Delta t$ becomes smaller. We therefore have a subtle order-of-limits issue, which can be analyzed by considering the effective temperature of the NESS as we send both $\Delta t \to 0$ and $\delta \to 1$ while keeping their ratio $r=\Delta t / (1-\delta)$ constant and finite. By recasting the problem in terms of variables $r$ and $\Delta t$ we find numerically that substantial deviations $\Delta \beta$ survive the $\Delta t\to 0$ limit for fixed values of $r>1$. However, we were not able to extract the non-Markovian analogue of Eq.~\eqref{no109819uj101u9} analytically. Our analysis shows that non-Markovian interactions have a significant effect on the dynamics in setting II, leading to a NESS with with a temperature different from the canonical temperature, even as $\Delta t\to 0$.

\acknowledgements
RMcE acknowledges support from Universities Ireland through the North/South Postgraduate Scholarship. RMcE and AKM acknowledge funding from Research Ireland through grant 21/RP2TF/10019. This work was supported by the John Templeton Foundation (Grant IDs 62422 and 63626).\\

\noindent
The data associated to this work is available from~\cite{McElvogue_NM-CM-SS_Code_2026}.

\bibliography{bibliography}

@article{Garg2025simulating,
  title = {Simulating quantum collision models with Hamiltonian simulations using early fault-tolerant quantum computers},
  author = {Garg, Kushagra and Ahmed, Zeeshan and Mitra, Subhadip and Chakraborty, Shantanav},
  journal = {Phys. Rev. A},
  volume = {112},
  issue = {2},
  pages = {022425},
  numpages = {25},
  year = {2025},
  month = {Aug},
  publisher = {American Physical Society},
  doi = {10.1103/3trk-smbh},
  url = {https://link.aps.org/doi/10.1103/3trk-smbh}
}

@article{Erbanni2023simulating,
  title = {Simulating quantum transport via collisional models on a digital quantum computer},
  author = {Erbanni, Rebecca and Xu, Xiansong and Demarie, Tommaso F. and Poletti, Dario},
  journal = {Phys. Rev. A},
  volume = {108},
  issue = {3},
  pages = {032619},
  numpages = {10},
  year = {2023},
  month = {Sep},
  publisher = {American Physical Society},
  doi = {10.1103/PhysRevA.108.032619},
  url = {https://link.aps.org/doi/10.1103/PhysRevA.108.032619}
}

@article{Cattaneo2021collision,
  title = {Collision Models Can Efficiently Simulate Any Multipartite Markovian Quantum Dynamics},
  author = {Cattaneo, Marco and De Chiara, Gabriele and Maniscalco, Sabrina and Zambrini, Roberta and Giorgi, Gian Luca},
  journal = {Phys. Rev. Lett.},
  volume = {126},
  issue = {13},
  pages = {130403},
  numpages = {8},
  year = {2021},
  month = {Apr},
  publisher = {American Physical Society},
  doi = {10.1103/PhysRevLett.126.130403},
  url = {https://link.aps.org/doi/10.1103/PhysRevLett.126.130403}
}

@article{cilluffo2021microscopic,
  title={Microscopic biasing of discrete-time quantum trajectories},
  author={Cilluffo, Dario and Buonaiuto, Giuseppe and Lesanovsky, Igor and Carollo, Angelo and Lorenzo, Salvatore and Palma, G Massimo and Ciccarello, Francesco and Carollo, Federico},
  journal={Quantum Science and Technology},
  volume={6},
  number={4},
  pages={045011},
  year={2021},
  publisher={IOP Publishing},
  url={https://doi.org/10.1088/2058-9565/ac15e2},
  doi={10.1088/2058-9565/ac15e2}
}

@article{cusumano2018interferometric,
  title = {Interferometric modulation of quantum cascade interactions},
  author = {Cusumano, Stefano and Mari, Andrea and Giovannetti, Vittorio},
  journal = {Phys. Rev. A},
  volume = {97},
  issue = {5},
  pages = {053811},
  numpages = {11},
  year = {2018},
  month = {May},
  publisher = {American Physical Society},
  doi = {10.1103/PhysRevA.97.053811},
  url = {https://link.aps.org/doi/10.1103/PhysRevA.97.053811}
}

@article{cusumano2017interferometric,
  title = {Interferometric quantum cascade systems},
  author = {Cusumano, Stefano and Mari, Andrea and Giovannetti, Vittorio},
  journal = {Phys. Rev. A},
  volume = {95},
  issue = {5},
  pages = {053838},
  numpages = {15},
  year = {2017},
  month = {May},
  publisher = {American Physical Society},
  doi = {10.1103/PhysRevA.95.053838},
  url = {https://link.aps.org/doi/10.1103/PhysRevA.95.053838}
}

@article{Lorenzo2015heat,
  title = {Heat flux and quantum correlations in dissipative cascaded systems},
  author = {Lorenzo, Salvatore and Farace, Alessandro and Ciccarello, Francesco and Palma, G. Massimo and Giovannetti, Vittorio},
  journal = {Phys. Rev. A},
  volume = {91},
  issue = {2},
  pages = {022121},
  numpages = {16},
  year = {2015},
  month = {Feb},
  publisher = {American Physical Society},
  doi = {10.1103/PhysRevA.91.022121},
  url = {https://link.aps.org/doi/10.1103/PhysRevA.91.022121}
}

@article{corr2025continuous,
  title={Continuous Variable Structured Collision Models},
  author={Corr, Anton and Cusumano, Stefano and De Chiara, Gabriele},
  journal={Quantum Science and Technology},
  volume={10},
  pages={045056},
  year={2025},
  url={https://doi.org/10.1088/2058-9565/ae0a7b},
  doi={10.1088/2058-9565/ae0a7b}
}

@article{hammam2021optimizing,
  title={Optimizing autonomous thermal machines powered by energetic coherence},
  author={Hammam, Kenza and Hassouni, Yassine and Fazio, Rosario and Manzano, Gonzalo},
  journal={New Journal of Physics},
  volume={23},
  number={4},
  pages={043024},
  year={2021},
  publisher={IOP Publishing},
  url={https://doi.org/10.1088/1367-2630/abeb47},
  doi={10.1088/1367-2630/abeb47}
}

@article{Hammam2024quantum,
  title = {Quantum coherence enables hybrid multitask and multisource regimes in autonomous thermal machines},
  author = {Hammam, Kenza and Manzano, Gonzalo and De Chiara, Gabriele},
  journal = {Phys. Rev. Res.},
  volume = {6},
  issue = {1},
  pages = {013310},
  numpages = {17},
  year = {2024},
  month = {Mar},
  publisher = {American Physical Society},
  doi = {10.1103/PhysRevResearch.6.013310},
  url = {https://link.aps.org/doi/10.1103/PhysRevResearch.6.013310}
}

@article{pedram2022environment,
  title={Environment-assisted modulation of heat flux in a bio-inspired system based on collision model},
  author={Pedram, Ali and {\c{C}}akmak, Bar{\i}{\c{s}} and M{\"u}stecapl{\i}o{\u{g}}lu, {\"O}zg{\"u}r E},
  journal={Entropy},
  volume={24},
  number={8},
  pages={1162},
  year={2022},
  publisher={MDPI},
  url={https://doi.org/10.3390/e24081162},
  doi={10.3390/e24081162}
}

@article{sabin2008classification,
  title={A classification of entanglement in three-qubit systems},
  author={Sab{\'\i}n, Carlos and Garc{\'\i}a-Alcaine, Guillermo},
  journal={The european physical journal D},
  volume={48},
  number={3},
  pages={435--442},
  year={2008},
  publisher={Springer},
  url={https://doi.org/10.1140/epjd/e2008-00112-5},
  doi={10.1140/epjd/e2008-00112-5}
}

@article{cusumano2022quantum,
  title={Quantum collision models: A beginner guide},
  author={Cusumano, S.},
  journal={Entropy},
  volume={24},
  number={9},
  pages={1258},
  year={2022},
  publisher={MDPI},
  url = {https://doi.org/10.3390/e24091258},
  doi = {10.3390/e24091258}
}

@article{cusumano2024structured,
  title={Structured quantum collision models: generating coherence with thermal resources},
  author={Cusumano, S. and De Chiara, G.},
  journal={New J. Phys.},
  volume={26},
  number={2},
  pages={023001},
  year={2024},
  publisher={IOP Publishing},
  doi = {10.1088/1367-2630/ad202a},
  url = {https://doi.org/10.1088/1367-2630/ad202a}
}

@article{karpat2025transient,
  title={Transient dynamics and homogenization in incoherent collision models},
  author={Karpat, G. and {\c{C}}akmak, B.},
  journal={Entropy},
  volume={27},
  number={2},
  pages={206},
  year={2025},
  publisher={MDPI}, 
  url = {https://doi.org/10.3390/e27020206},
  doi = {10.3390/e27020206}
}

@article{csenyacsa2022entropy,
  title={Entropy Production in Non-Markovian Collision Models: Information Backflow vs. System-Environment Correlations},
  author={{\c{S}}enya{\c{s}}a, H. T. and Kesgin, {\c{S}}. and Karpat, G. and {\c{C}}akmak, B.},
  journal={Entropy},
  volume={24},
  number={6},
  pages={824},
  year={2022},
  publisher={Mdpi},
  url ={https://doi.org/10.3390/e24060824},
  doi = {10.3390/e24060824}
}

@article{karpat2019pra,
  title = {Quantum synchronization in a collision model},
  author = {Karpat, G. and Yal{\c{c}}{\i}nkaya, I. and {\c{C}}akmak, B.},
  journal = {Phys. Rev. A},
  volume = {100},
  issue = {1},
  pages = {012133},
  numpages = {11},
  year = {2019},
  month = {Jul},
  publisher = {American Physical Society},
  doi = {10.1103/PhysRevA.100.012133},
  url = {https://link.aps.org/doi/10.1103/PhysRevA.100.012133}
}

@article{Deutsch2018, 
title={Eigenstate thermalization hypothesis}, 
volume={81}, 
url={http://dx.doi.org/10.1088/1361-6633/aac9f1}, 
DOI={10.1088/1361-6633/aac9f1}, 
journal={Rep. Prog. Phys.},
author={Deutsch, J. M.}, 
year={2018}, 
pages={082001} 
}

@article{goldhaber1998kondo,
  title={Kondo effect in a single-electron transistor},
  author={Goldhaber-Gordon, David and Shtrikman, Hadas and Mahalu, Diana and Abusch-Magder, David and Meirav, U and Kastner, MA},
  journal={Nature},
  volume={391},
  number={6663},
  pages={156--159},
  year={1998},
  publisher={Nature Publishing Group UK London},
  url={https://doi.org/10.1038/34373} 
}

@article{shim2018numerical,
  title={Numerical renormalization group method for entanglement negativity at finite temperature},
  author={Shim, Jeongmin and Sim, H-S and Lee, Seung-Sup B},
  journal={Physical Review B},
  volume={97},
  number={15},
  pages={155123},
  year={2018},
  publisher={APS},
  url={https://doi.org/10.1103/PhysRevB.97.155123} 
}

@article{barthelemy2013quantum,
  title={Quantum dot systems: a versatile platform for quantum simulations},
  author={Barthelemy, Pierre and Vandersypen, Lieven MK},
  journal={Annalen der Physik},
  volume={525},
  number={10-11},
  pages={808--826},
  year={2013},
  publisher={Wiley Online Library},
  url={https://doi.org/10.1002/andp.201300124}
}

@article{iftikhar2018tunable,
  title={Tunable quantum criticality and super-ballistic transport in a “charge” Kondo circuit},
  author={Iftikhar, Z and Anthore, A and Mitchell, AK and Parmentier, FD and Gennser, U and Ouerghi, A and Cavanna, A and Mora, C and Simon, P and Pierre, F},
  journal={Science},
  volume={360},
  number={6395},
  pages={1315--1320},
  year={2018},
  publisher={American Association for the Advancement of Science},
  url={https://doi.org/10.1126/science.aan5592}
}

@article{pouse2023quantum,
  title={Quantum simulation of an exotic quantum critical point in a two-site charge Kondo circuit},
  author={Pouse, Winston and Peeters, Lucas and Hsueh, Connie L and Gennser, Ulf and Cavanna, Antonella and Kastner, Marc A and Mitchell, Andrew K and Goldhaber-Gordon, David},
  journal={Nature Physics},
  volume={19},
  number={4},
  pages={492--499},
  year={2023},
  publisher={Nature Publishing Group UK London},
  url={https://doi.org/10.1038/s41567-022-01905-4}
}

@article{petropoulos2024nanoscale,
  title={Nanoscale single-electron box with a floating lead for quantum sensing: Modeling and device characterization},
  author={Petropoulos, Nikolaos and Wu, Xutong and Sokolov, Andrii and Giounanlis, Panagiotis and Bashir, Imran and Mitchell, Andrew K and Asker, Mike and Leipold, Dirk and Staszewski, Robert B and Blokhina, Elena},
  journal={Applied Physics Letters},
  volume={124},
  number={17},
  year={2024},
  publisher={AIP Publishing},
  url={https://doi.org/10.1063/5.0203421}
}

@misc{ManiscalcoScalability,
      title={Scalability of quantum error mitigation techniques: from utility to advantage}, 
      author={S. N. Filippov and S. Maniscalco and Guillermo Garc\'ia-P\'erez},
      year={2024},
      eprint={2403.13542},
      archivePrefix={arXiv},
      primaryClass={quant-ph},
      url={https://arxiv.org/abs/2403.13542} 
}

@book{Breuer2002,
author = {Breuer, H.-P. and Petruccione, F.},
doi = {10.1093/acprof:oso/9780199213900.001.0001},
publisher = {Oxford University Press},
title = {{The Theory of Open Quantum Systems}},
year = {2002}
}

@article{GooldPRX,
  title = {Tensor-Network Method to Simulate Strongly Interacting Quantum Thermal Machines},
  author = {Brenes, M. and Mendoza-Arenas, J. J. and Purkayastha, A. and Mitchison, M. T. and Clark, S. R. and Goold, J.},
  journal = {Phys. Rev. X},
  volume = {10},
  issue = {3},
  pages = {031040},
  numpages = {29},
  year = {2020},
  month = {Aug},
  publisher = {American Physical Society},
  doi = {10.1103/PhysRevX.10.031040},
  url = {https://link.aps.org/doi/10.1103/PhysRevX.10.031040}
}

@article{HuberCooling,
  title = {Exponential Improvement for Quantum Cooling through Finite-Memory Effects},
  author = {Taranto, P. and Bakhshinezhad, F. and Sch\"uttelkopf, P. and Clivaz, F. and Huber, M.},
  journal = {Phys. Rev. Appl.},
  volume = {14},
  issue = {5},
  pages = {054005},
  numpages = {20},
  year = {2020},
  month = {Nov},
  publisher = {American Physical Society},
  doi = {10.1103/PhysRevApplied.14.054005},
  url = {https://link.aps.org/doi/10.1103/PhysRevApplied.14.054005}
}

@misc{KochCooling,
      title={Universal cooling of quantum systems via randomized measurements}, 
      author={J. Langbehn and G. Mouloudakis and E. King and R. Menu and I. Gornyi and G. Morigi and Y. Gefen and C. P. Koch},
      year={2025},
      eprint={2506.11964},
      archivePrefix={arXiv},
      primaryClass={quant-ph},
      url={https://arxiv.org/abs/2506.11964} 
}

@article{LandiRMP,
  title = {Nonequilibrium boundary-driven quantum systems: Models, methods, and properties},
  author = {Landi, G. T. and Poletti, D. and Schaller, G.},
  journal = {Rev. Mod. Phys.},
  volume = {94},
  issue = {4},
  pages = {045006},
  numpages = {58},
  year = {2022},
  month = {Dec},
  publisher = {American Physical Society},
  doi = {10.1103/RevModPhys.94.045006},
  url = {https://link.aps.org/doi/10.1103/RevModPhys.94.045006}
}

@article{Scattering1, 
title={Micro-reversibility and thermalization with collisional baths}, volume={552}, 
url={http://dx.doi.org/10.1016/j.physa.2019.122108}, 
DOI={10.1016/j.physa.2019.122108}, 
journal={Physica A}, 
author={Ehrich, J. and Esposito, M. and Barra, F. and Parrondo, J. M. R.}, 
year={2020}, 
pages={122108} 
}

@article{Scattering2,
  title = {Thermalization Induced by Quantum Scattering},
  author = {Jacob, S. L. and Esposito, M. and Parrondo, J. M.R. and Barra, F.},
  journal = {PRX Quantum},
  volume = {2},
  issue = {2},
  pages = {020312},
  numpages = {19},
  year = {2021},
  month = {Apr},
  publisher = {American Physical Society},
  doi = {10.1103/PRXQuantum.2.020312},
  url = {https://link.aps.org/doi/10.1103/PRXQuantum.2.020312}
}

@article{Scattering3,
  title = {Thermalization and Dephasing in Collisional Reservoirs},
  author = {Tabanera-Bravo, J. and Parrondo, J. M. R. and Esposito, M. and Barra, F.},
  journal = {Phys. Rev. Lett.},
  volume = {130},
  issue = {20},
  pages = {200402},
  numpages = {6},
  year = {2023},
  month = {May},
  publisher = {American Physical Society},
  doi = {10.1103/PhysRevLett.130.200402},
  url = {https://link.aps.org/doi/10.1103/PhysRevLett.130.200402}
}

@article{PReB,
  title = {Periodically refreshed baths to simulate open quantum many-body dynamics},
  author = {Purkayastha, A. and Guarnieri, G. and Campbell, S. and Prior, J. and Goold, J.},
  journal = {Phys. Rev. B},
  volume = {104},
  issue = {4},
  pages = {045417},
  numpages = {16},
  year = {2021},
  month = {Jul},
  publisher = {American Physical Society},
  doi = {10.1103/PhysRevB.104.045417},
  url = {https://link.aps.org/doi/10.1103/PhysRevB.104.045417}
}

@article{StrasbergPRX,
  title = {Quantum and Information Thermodynamics: A Unifying Framework Based on Repeated Interactions},
  author = {Strasberg, P. and Schaller, G. and Brandes, T. and Esposito, M.},
  journal = {Phys. Rev. X},
  volume = {7},
  issue = {2},
  pages = {021003},
  numpages = {33},
  year = {2017},
  month = {Apr},
  publisher = {American Physical Society},
  doi = {10.1103/PhysRevX.7.021003},
  url = {https://link.aps.org/doi/10.1103/PhysRevX.7.021003}
}

@article{BarisPRA2017,
  title = {Non-Markovianity, coherence, and system-environment correlations in a long-range collision model},
  author = {\c{C}akmak, B. and Pezzutto, M. and Paternostro, M. and M\"ustecapl\i o\u{g}lu, \"O. E.},
  journal = {Phys. Rev. A},
  volume = {96},
  issue = {2},
  pages = {022109},
  numpages = {8},
  year = {2017},
  month = {Aug},
  publisher = {American Physical Society},
  doi = {10.1103/PhysRevA.96.022109},
  url = {https://link.aps.org/doi/10.1103/PhysRevA.96.022109}
}

@article{CampbellPRA2019,
  title = {Collisional unfolding of quantum Darwinism},
  author = {Campbell, S. and \c{C}akmak, B. and M\"ustecapl\i o\u{g}lu, \"O. E. and Paternostro, M. and Vacchini, B.},
  journal = {Phys. Rev. A},
  volume = {99},
  issue = {4},
  pages = {042103},
  numpages = {7},
  year = {2019},
  month = {Apr},
  publisher = {American Physical Society},
  doi = {10.1103/PhysRevA.99.042103},
  url = {https://link.aps.org/doi/10.1103/PhysRevA.99.042103}
}

@article{Ryan2022, 
title={Commutativity and the emergence of classical objectivity}, 
volume={6}, 
url={http://dx.doi.org/10.1088/2399-6528/ac8f19}, 
DOI={10.1088/2399-6528/ac8f19}, 
journal={J. Phys. Commun.}, 
author={Ryan, E. and Carolan, E. and Campbell, S. and Paternostro, M.}, year={2022}, 
pages={095005}
}

@article{ZurekRMP,
  title = {Decoherence, einselection, and the quantum origins of the classical},
  author = {Zurek, W. H.},
  journal = {Rev. Mod. Phys.},
  volume = {75},
  issue = {3},
  pages = {715--775},
  numpages = {0},
  year = {2003},
  month = {May},
  publisher = {American Physical Society},
  doi = {10.1103/RevModPhys.75.715},
  url = {https://link.aps.org/doi/10.1103/RevModPhys.75.715}
}

@article{Zhang2019, 
title={Non-Markovianity in the presence of multiple thermal environments via collision model}, 
volume={383}, 
url={http://dx.doi.org/10.1016/j.physleta.2019.05.002}, 
DOI={10.1016/j.physleta.2019.05.002}, 
journal={Phys. Lett. A}, 
author={Zhang, Q. and Man, Z.-X. and Xia, Y.-J.}, 
year={2019}, 
pages={2456–2461} 
}

@article{Baris1,
  title = {Heat transport and rectification via quantum statistical and coherence asymmetries},
  author = {Palafox, S. and Rom\'an-Ancheyta, R. and \c{C}akmak, B. and M\"ustecapl{\i}o{\u{g}}lu, \"O. E.},
  journal = {Phys. Rev. E},
  volume = {106},
  issue = {5},
  pages = {054114},
  numpages = {10},
  year = {2022},
  month = {Nov},
  publisher = {American Physical Society},
  doi = {10.1103/PhysRevE.106.054114},
  url = {https://link.aps.org/doi/10.1103/PhysRevE.106.054114}
}

@article{Baris2, 
title={Quantum thermal machine as a rectifier}, 
volume={10},
url={http://dx.doi.org/10.1088/2058-9565/adaee0}, 
DOI={10.1088/2058-9565/adaee0}, 
journal={Quantum Sci. Technol.}, 
author={Santiago-García, M. and Pusuluk, O. and M\"ustecapl{\i}o\u{g}lu, \"O. E. and \c{C}akmak, B. and Rom\'an-Ancheyta, R.}, year={2025}, 
pages={025018} 
}

@article{ScaraniPRL2002,
  title = {Thermalizing Quantum Machines: Dissipation and Entanglement},
  author = {Scarani, V. and Ziman, M. and \v{S}telmachovi\v{c}, P. and Gisin, N. and Bu\v{z}ek, V.},
  journal = {Phys. Rev. Lett.},
  volume = {88},
  issue = {9},
  pages = {097905},
  numpages = {4},
  year = {2002},
  month = {Feb},
  publisher = {American Physical Society},
  doi = {10.1103/PhysRevLett.88.097905},
  url = {https://link.aps.org/doi/10.1103/PhysRevLett.88.097905}
}

@article{CICCARELLO20221,
title = {Quantum collision models: Open system dynamics from repeated interactions},
journal = {Physics Reports},
volume = {954},
pages = {1-70},
year = {2022},
doi = {https://doi.org/10.1016/j.physrep.2022.01.001},
url = {https://www.sciencedirect.com/science/article/pii/S0370157322000035},
author = {F. Ciccarello and S. Lorenzo and V. Giovannetti and G. M. Palma},
}

@article{GiovannettiPRA2018,
  title = {Entropy production and asymptotic factorization via thermalization: A collisional model approach},
  author = {Cusumano, S. and Cavina, V. and Keck, M. and De Pasquale, A. and Giovannetti, V.},
  journal = {Phys. Rev. A},
  volume = {98},
  issue = {3},
  pages = {032119},
  numpages = {9},
  year = {2018},
  month = {Sep},
  publisher = {American Physical Society},
  doi = {10.1103/PhysRevA.98.032119},
  url = {https://link.aps.org/doi/10.1103/PhysRevA.98.032119}
}

@article{Guarnieri2020PLA, 
title={Non-equilibrium steady-states of memoryless quantum collision models}, 
volume={384}, 
url={http://dx.doi.org/10.1016/j.physleta.2020.126576}, 
DOI={10.1016/j.physleta.2020.126576}, 
journal={Phys. Lett. A}, 
author={Guarnieri, G. and Morrone, D. and \c{C}akmak, B. and Plastina, F. and Campbell, S.}, 
year={2020}, 
pages={126576} 
}

@article{Segal2025QST, 
title={Equilibrium and nonequilibrium steady states with the repeated interaction protocol: relaxation dynamics and energetic cost}, 
volume={10}, 
url={http://dx.doi.org/10.1088/2058-9565/adc7d4}, 
DOI={10.1088/2058-9565/adc7d4}, 
journal={Quantum Sci. Technol.}, 
author={Prositto, A. and Forbes, M. and Segal, D.}, 
year={2025}, 
pages={025061} 
}

@article{prositto2025collisional,
  title={Collisional model with dissipative and dephasing baths: Nonadditive effects at strong coupling},
  author={Prositto, Alessandro and Ramon-Escandell, Carlos and Segal, Dvira},
  journal={arXiv preprint arXiv:2509.10988},
  year={2025},
  doi={10.48550/arXiv.2509.10988},
  url={https://doi.org/10.48550/arXiv.2509.10988}
}

@article{Current_fluctuations,
  title = {Current Fluctuations in Open Quantum Systems: Bridging the Gap Between Quantum Continuous Measurements and Full Counting Statistics},
  author = {Landi, G. T. and Kewming, M. J. and Mitchison, M. T. and Potts, P. P.},
  journal = {PRX Quantum},
  volume = {5},
  issue = {2},
  pages = {020201},
  numpages = {86},
  year = {2024},
  month = {Apr},
  publisher = {American Physical Society},
  doi = {10.1103/PRXQuantum.5.020201},
  url = {https://link.aps.org/doi/10.1103/PRXQuantum.5.020201}
}

@article{2-level_thermalization,
doi = {10.1209/0295-5075/126/40003},
url = {https://dx.doi.org/10.1209/0295-5075/126/40003},
year = {2019},
volume = {126},
number = {4},
pages = {40003},
author = {J. P. Cherian and S. Chakraborty and S. Ghosh},
title = {On thermalization of two-level quantum systems},
journal = {EPL},
}

@article{Bassano,
  title = {System-environment correlations and Markovian embedding of quantum non-Markovian dynamics},
  author = {Campbell, S. and Ciccarello, F. and Palma, G. M. and Vacchini, B.},
  journal = {Phys. Rev. A},
  volume = {98},
  issue = {1},
  pages = {012142},
  numpages = {11},
  year = {2018},
  month = {Jul},
  publisher = {American Physical Society},
  doi = {10.1103/PhysRevA.98.012142},
  url = {https://link.aps.org/doi/10.1103/PhysRevA.98.012142}
}

@article{Campbell2021EPL, 
title={Collision models in open system dynamics: A versatile tool for deeper insights?}, 
volume={133},
url={http://dx.doi.org/10.1209/0295-5075/133/60001}, 
DOI={10.1209/0295-5075/133/60001}, 
journal={EPL}, 
author={Campbell, S. and Vacchini, B.},
year={2021},
pages={60001} 
}

@article{Rau1963,
  title = {Relaxation Phenomena in Spin and Harmonic Oscillator Systems},
  author = {Rau, J.},
  journal = {Phys. Rev.},
  volume = {129},
  issue = {4},
  pages = {1880--1888},
  numpages = {0},
  year = {1963},
  month = {Feb},
  publisher = {American Physical Society},
  doi = {10.1103/PhysRev.129.1880},
  url = {https://link.aps.org/doi/10.1103/PhysRev.129.1880}
}

@article{ruariMauro,
  title = {Non-Markovianity and system-environment correlations in a microscopic collision model},
  author = {McCloskey, R. and Paternostro, M.},
  journal = {Phys. Rev. A},
  volume = {89},
  issue = {5},
  pages = {052120},
  numpages = {6},
  year = {2014},
  month = {May},
  publisher = {American Physical Society},
  doi = {10.1103/PhysRevA.89.052120},
  url = {https://link.aps.org/doi/10.1103/PhysRevA.89.052120}
}

@article{NMmeasures,
  title = {Comparative study of non-Markovianity measures in exactly solvable one- and two-qubit models},
  author = {Addis, C. and Bylicka, B. and Chru\'{s}ci\'{n}ski, D. and Maniscalco, S.},
  journal = {Phys. Rev. A},
  volume = {90},
  issue = {5},
  pages = {052103},
  numpages = {17},
  year = {2014},
  month = {Nov},
  publisher = {American Physical Society},
  doi = {10.1103/PhysRevA.90.052103},
  url = {https://link.aps.org/doi/10.1103/PhysRevA.90.052103}
}

@article{BLPmeasure,
  title = {Measure for the Degree of Non-Markovian Behavior of Quantum Processes in Open Systems},
  author = {Breuer, H.-P. and Laine, E.-M. and Piilo, J.},
  journal = {Phys. Rev. Lett.},
  volume = {103},
  issue = {21},
  pages = {210401},
  numpages = {4},
  year = {2009},
  month = {Nov},
  publisher = {American Physical Society},
  doi = {10.1103/PhysRevLett.103.210401},
  url = {https://link.aps.org/doi/10.1103/PhysRevLett.103.210401}
}

@article{OptimalNM_states,
  title = {Optimal state pairs for non-Markovian quantum dynamics},
  author = {Wi\ss{}mann, S. and Karlsson, A. and Laine, E.-M. and Piilo, J. and Breuer, H.-P.},
  journal = {Phys. Rev. A},
  volume = {86},
  issue = {6},
  pages = {062108},
  numpages = {6},
  year = {2012},
  month = {Dec},
  publisher = {American Physical Society},
  doi = {10.1103/PhysRevA.86.062108},
  url = {https://link.aps.org/doi/10.1103/PhysRevA.86.062108}
}

@article{Composite_CM,
  title = {Composite quantum collision models},
  author = {Lorenzo, S. and Ciccarello, F. and Palma, G. M.},
  journal = {Phys. Rev. A},
  volume = {96},
  issue = {3},
  pages = {032107},
  numpages = {11},
  year = {2017},
  month = {Sep},
  publisher = {American Physical Society},
  doi = {10.1103/PhysRevA.96.032107},
  url = {https://link.aps.org/doi/10.1103/PhysRevA.96.032107}
}

@article{AA_to_Composite,
  title = {Collision model for non-Markovian quantum dynamics},
  author = {Kretschmer, S. and Luoma, K. and Strunz, W. T.},
  journal = {Phys. Rev. A},
  volume = {94},
  issue = {1},
  pages = {012106},
  numpages = {9},
  year = {2016},
  month = {Jul},
  publisher = {American Physical Society},
  doi = {10.1103/PhysRevA.94.012106},
  url = {https://link.aps.org/doi/10.1103/PhysRevA.94.012106}
}

@article{MEsfromCMs,
author = {Ziman, M. and \v{S}telmachovi\v{c}, P. and Bu\v{z}ek, V.},
title = {Description of Quantum Dynamics of Open Systems Based on Collision-Like Models},
journal = {Open Syst. Inf. Dyn.},
volume = {12},
number = {01},
pages = {81-91},
year = {2005},
doi = {10.1007/s11080-005-0488-0},
URL = {https://doi.org/10.1007/s11080-005-0488-0}
}

@article{NmReview,
  title = {Colloquium: Non-Markovian dynamics in open quantum systems},
  author = {Breuer, H.-P. and Laine, E.-M. and Piilo, J. and Vacchini, B.},
  journal = {Rev. Mod. Phys.},
  volume = {88},
  issue = {2},
  pages = {021002},
  numpages = {24},
  year = {2016},
  month = {Apr},
  publisher = {American Physical Society},
  doi = {10.1103/RevModPhys.88.021002},
  url = {https://link.aps.org/doi/10.1103/RevModPhys.88.021002}
}

@article{BLPmeasure2,
  title = {Measure for the non-Markovianity of quantum processes},
  author = {Laine, E.-M. and Piilo, J. and Breuer, H.-P.},
  journal = {Phys. Rev. A},
  volume = {81},
  issue = {6},
  pages = {062115},
  numpages = {8},
  year = {2010},
  month = {Jun},
  publisher = {American Physical Society},
  doi = {10.1103/PhysRevA.81.062115},
  url = {https://link.aps.org/doi/10.1103/PhysRevA.81.062115}
}

@article{Homogenisation,
  title = {Diluting quantum information: An analysis of information transfer in system-reservoir interactions},
  author = {Ziman, M. and \ifmmode \check{S}\else \v{S}\fi{}telmachovi\ifmmode \check{c}\else \v{c}\fi{}, P. and Bu\ifmmode \check{z}\else \v{z}\fi{}ek, V. and Hillery, M. and Scarani, V. and Gisin, N.},
  journal = {Phys. Rev. A},
  volume = {65},
  issue = {4},
  pages = {042105},
  numpages = {11},
  year = {2002},
  month = {Mar},
  publisher = {American Physical Society},
  doi = {10.1103/PhysRevA.65.042105},
  url = {https://link.aps.org/doi/10.1103/PhysRevA.65.042105}
}

@article{BiPartite_Neg,
  title = {Separability Criterion for Density Matrices},
  author = {Peres, A.},
  journal = {Phys. Rev. Lett.},
  volume = {77},
  issue = {8},
  pages = {1413--1415},
  numpages = {0},
  year = {1996},
  month = {Aug},
  publisher = {American Physical Society},
  doi = {10.1103/PhysRevLett.77.1413},
  url = {https://link.aps.org/doi/10.1103/PhysRevLett.77.1413}
}

@article{Ferracin2024, 
title={Spectral density modulation and universal Markovian closure of fermionic environments}, 
volume={161}, 
url={http://dx.doi.org/10.1063/5.0226723}, 
DOI={10.1063/5.0226723}, 
journal={J. Chem. Phys.},
author={Ferracin, D. and Smirne, A. and Huelga, S. F. and Plenio, M. B. and Tamascelli, D.}, 
year={2024},
pages={174114}
}

@article{Heineken_NESS_Entanglement,
  title = {Quantum-memory-enhanced dissipative entanglement creation in nonequilibrium steady states},
  author = {Heineken, D. and Beyer, K. and Luoma, K. and Strunz, W. T.},
  journal = {Phys. Rev. A},
  volume = {104},
  issue = {5},
  pages = {052426},
  numpages = {11},
  year = {2021},
  month = {Nov},
  publisher = {American Physical Society},
  doi = {10.1103/PhysRevA.104.052426},
  url = {https://link.aps.org/doi/10.1103/PhysRevA.104.052426}
}

@article{Tian_Heat_Transfer_2021,
	title = {Effect of Inter-System Coupling on Heat Transport in a Microscopic Collision Model},
	volume = {23},
	copyright = {http://creativecommons.org/licenses/by/3.0/},
	issn = {1099-4300},
	url = {https://www.mdpi.com/1099-4300/23/4/471},
	doi = {10.3390/e23040471},
	number = {4},
	urldate = {2025-05-06},
	journal = {Entropy},
	author = {Tian, F. and Zou, J. and Li, L. and Li, H. and Shao, B.},
	month = {Apr},
	year = {2021},
	pages = {471},
}

@article{Pezzutto_Paternostro_2016,
doi = {10.1088/1367-2630/18/12/123018},
url = {https://dx.doi.org/10.1088/1367-2630/18/12/123018},
year = {2016},
month = {dec},
publisher = {IOP Publishing},
volume = {18},
number = {12},
pages = {123018},
author = {Pezzutto, M. and Paternostro, M. and Omar, Y.},
title = {Implications of non-Markovian quantum dynamics for the Landauer bound},
journal = {New Journal of Physics}
}

@misc{Pleasance_2025,
      title={Non-Markovianity in collision models with initial intra-environment correlations}, 
      author={G. Pleasance and A. E. Neira and M. Merkli and F. Petruccione},
      year={2025},
      eprint={2505.05433},
      archivePrefix={arXiv},
      primaryClass={quant-ph},
      url={https://arxiv.org/abs/2505.05433}, 
}

@article{DeChiaraNJP, 
title={Reconciliation of quantum local master equations with thermodynamics}, 
volume={20}, 
url={http://dx.doi.org/10.1088/1367-2630/aaecee}, 
DOI={10.1088/1367-2630/aaecee}, 
number={11}, 
journal={New J. Phys.}, 
author={De Chiara, G. and Landi, G. and Hewgill, A. and Reid, B. and Ferraro, A. and Roncaglia, A. J. and Antezza, M.}, 
year={2018}, 
pages={113024} 
}

@article{Brun_CM_Trajectories,
    author = {Brun, T. A.},
    title = {A simple model of quantum trajectories},
    journal = {American Journal of Physics},
    volume = {70},
    number = {7},
    pages = {719-737},
    year = {2002},
    month = {07},
    issn = {0002-9505},
    doi = {10.1119/1.1475328},
    url = {https://doi.org/10.1119/1.1475328},
}

@article{Whalen_NMCM_Trajectories,
  title = {Collision model for non-Markovian quantum trajectories},
  author = {Whalen, S. J.},
  journal = {Phys. Rev. A},
  volume = {100},
  issue = {5},
  pages = {052113},
  numpages = {7},
  year = {2019},
  month = {Nov},
  publisher = {American Physical Society},
  doi = {10.1103/PhysRevA.100.052113},
  url = {https://link.aps.org/doi/10.1103/PhysRevA.100.052113}
}

@article{Kretschmer_CompositeCM,
  title = {Collision model for non-Markovian quantum dynamics},
  author = {S. Kretschmer and K. Luoma and  W. T. Strunz},
  journal = {Phys. Rev. A},
  volume = {94},
  issue = {1},
  pages = {012106},
  numpages = {9},
  year = {2016},
  month = {Jul},
  publisher = {American Physical Society},
  doi = {10.1103/PhysRevA.94.012106},
  url = {https://link.aps.org/doi/10.1103/PhysRevA.94.012106}
}

@misc{Lacroix2024MakingQCMExact,
      title={Making Quantum Collision Models Exact}, 
      author={T. Lacroix and D. Cilluffo and S. F. Huelga and M. B. Plenio},
      year={2024},
      eprint={2411.13166},
      archivePrefix={arXiv},
      primaryClass={quant-ph},
      url={https://arxiv.org/abs/2411.13166}, 
}

@article{HarwoodCMFeedback,
  title = {Unified collision model of coherent and measurement-based quantum feedback},
  author = {Harwood, A. and Brunelli, M. and Serafini, A.},
  journal = {Phys. Rev. A},
  volume = {108},
  issue = {4},
  pages = {042413},
  numpages = {17},
  year = {2023},
  month = {Oct},
  publisher = {American Physical Society},
  doi = {10.1103/PhysRevA.108.042413},
  url = {https://link.aps.org/doi/10.1103/PhysRevA.108.042413}
}

@software{McElvogue_NM-CM-SS_Code_2026,
author = {McElvogue, R.},
month = {jan},
title = {https://github.com/ronanmcelvogue/Non-Markovian-CM-SS},
url = {https://github.com/ronanmcelvogue/Non-Markovian-CM-SS},
version = {1.0.0},
year = {2026}
}

\end{document}